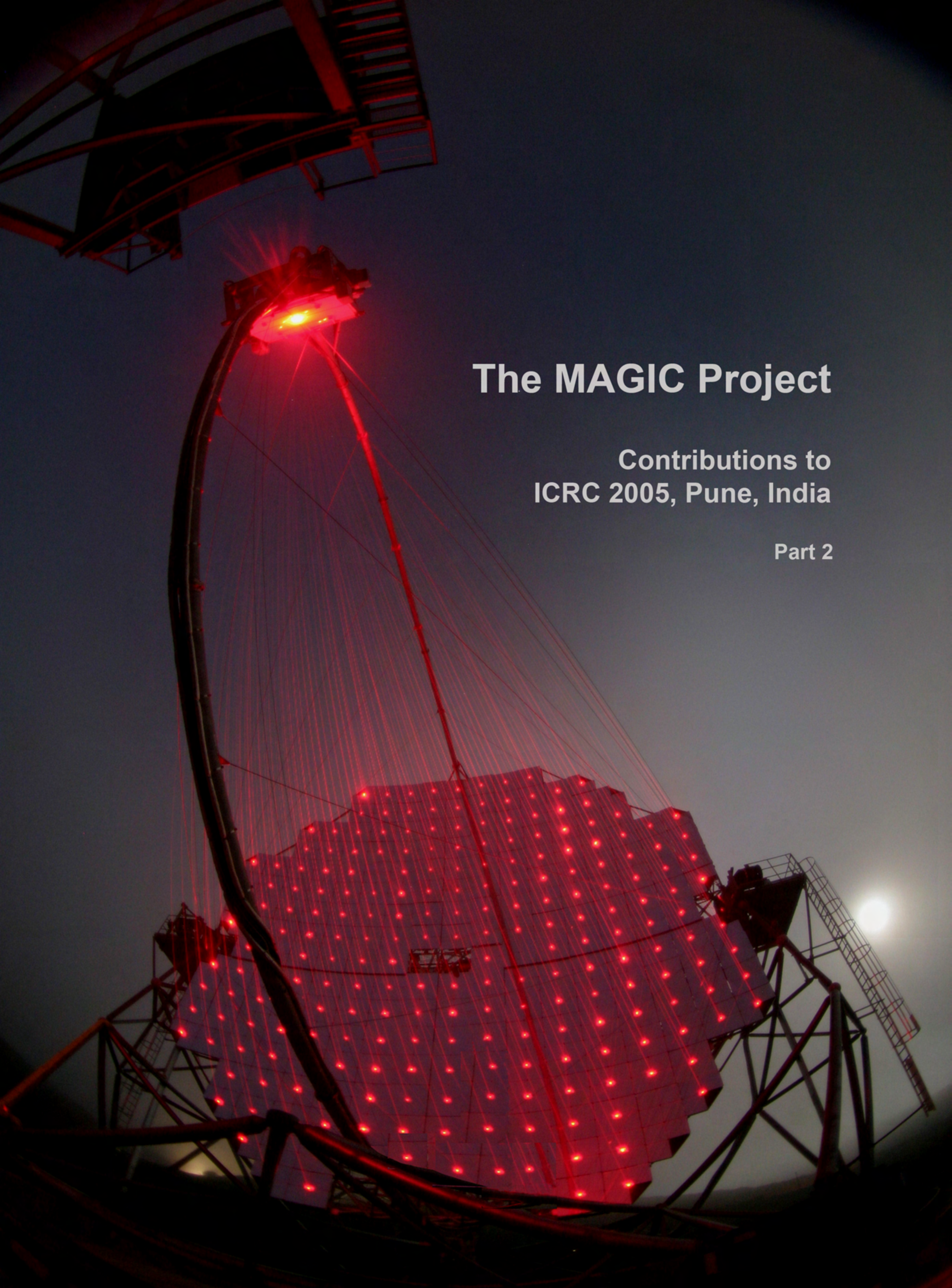

# The MAGIC Project

## Contributions to ICRC 2005, Pune, India

### Part 2

Contents

**Contributions:**

**Part 2: Future Plans and Developments**

# MAGIC: List of collaboration members

J. Albert i Fort [a], E. Aliu [b], H. Anderhub [g], P. Antoranz [k], A. Armada [b], M. Asensio [k], C. Baixeras [c], J.A. Barrio [k], H. Bartko [d], D. Bastieri [e], W. Bednarek [f], K. Berger [a], C. Bigongiari [e], A. Biland [g], E. Bisesi [h], O. Blanch [b], R.K. Bock [d], T. Bretz [a], I. Britvitch [g], M. Camara [k], A. Chilingarian [i], S. Ciprini [j], J.A. Coarasa [d], S. Commichau [g], J.L. Contreras [k], J. Cortina [b], V. Danielyan [i], F. Dazzi [e], A. De Angelis [h], B. De Lotto [h], E. Domingo [b], D. Dorner [a], M. Doro [b], O. Epler [l], M. Errando [b], D. Ferenc [m], E. Fernandez [b], R. Firpo [b], J. Flix [b], M.V. Fonseca [k], L. Font [c], N. Galante [n], M. Garczarczyk [d], M. Gaug [b], J. Gebauer [d], R. Giannitrapani [h], M. Giller [f], F. Goebel [d], D. Hakobyan [i], M. Hayashida [d], T. Hengstebeck [l], D. Höhne [a], J. Hose [d], P. Jacon [f], O.C. de Jager [o], O. Kalekin [l], D. Kranich [m], A. Laille [m], T. Lenisa [h], P. Liebing [d], E. Lindfors [j], F. Longo [h], M. Lopez [k], J. Lopez [b], E. Lorenz [d, g], F. Lucarelli [k], P. Majumdar [d], G. Maneva [q], K. Mannheim [a], M. Mariotti [e], M. Martinez [b], K. Mase [d], D. Mazin [d], C. Merck [d], M. Merck [a], M. Meucci [n], M. Meyer [a], J.M. Miranda [k], R. Mirzoyan [d], S. Mizobuchi [d], A. Moralejo [e], E. Ona-Wilhelmi [b], R. Orduna [c], N. Otte [d], I.Oya [k], D. Paneque [d], R. Paoletti [n], M. Pasanen [j], D. Pascoli [e], F. Pauss [g], N. Pavel [l], R. Pegna [n], L. Peruzzo [e], A. Piccioli [n], M. Pin [h], E. Prandini [e], R. de los Reyes [k], J. Rico [b], W. Rhode [p], B. Riegel [a], M. Rissi [g], A. Robert [c], G. Rossato [e], S. Rügamer [a], A. Saggion [e], A. Sanchez [e], P. Sartori [e], V. Scalzotto [e], R. Schmitt [a], T. Schweizer [l], M. Shayduk [l], K. Shinozaki [d], N. Sidro [b], A. Sillanpää [j], D. Sobczynska [f], A. Stamerra [n], L. Stark [g], L. Takalo [j], P. Temnikov [q], D. Tescaro [e], M. Teshima [d], N. Tonello [d], A. Torres [c], N. Turini [n], H. Vankov [q], V. Vitale [d], S. Volkov [l], R. Wagner [d], T. Wibig [f], W. Wittek [d], J. Zapatero [c]

[a] Universität Würzburg, Germany
[b] Institut de Fisica d'Altes Energies, Barcelona, Spain
[c] Universitat Autonoma de Barcelona, Spain
[d] Max-Planck-Institut für Physik, München, Germany
[e] Dipartimento di Fisica, Università di Padova, and INFN Padova, Italy
[f] Division of Experimental Physics, University of Lodz, Poland
[g] Institute for Particle Physics, ETH Zürich, Switzerland
[h] Dipartimento di Fisica, Università di Udine, and INFN Trieste, Italy
[i] Yerevan Physics Institute, Cosmic Ray Division, Yerevan, Armenia
[j] Tuorla Observatory, Pikkiö, Finland
[k] Universidad Complutense, Madrid, Spain
[l] Institut für Physik, Humboldt-Universität Berlin, Germany
[m] University of California, Davis, USA
[n] Dipartimento di Fisica, Università di Siena, and INFN Pisa, Italy

[o] Space Research Unit, Potchefstroom University, South Africa
[p] Fachbereich Physik, Universität Dortmund, Germany
[q] Institute for Nuclear Research and Nuclear Energy, Sofia, Bulgaria

# Part 2: Future Plans and Developments

# Towards Dark Matter Searches with the MAGIC Telescope

H. Bartko[a], A. Biland[b], E. Bisesi[c], D. Elsässer[d], P. Flix[e], P. Häfliger[b],
M. Mariotti[f], S. Stark[b], W. Wittek[a] for the MAGIC collaboration

*(a) Max Planck Institute for Physics, Munich Germany*
*(b) ETH Zurich, Switzerland*
*(c) University of Udine and INFN Trieste, Italy*
*(d) University of Wüerzburg, Germany*
*(d) Institut de Fisica d Altes Energies, Edifici Cn Universitat Autonoma de Barcelona, Bellaterra, Spain*
*(e) University and INFN Padova, Italy.*

Presenter: H. Bartko (hbartko@mppmu.mpg.de), ger-bartko-H-abs3-og21-poster

MAGIC is a 17m diameter Imaging Air Cherenkov Telescope installed on the Canary Island La Palma. The telescope is designed for gamma-ray astronomy in the 30 GeV to 30 TeV energy range. Particle physics models predict candidate particles for Dark Matter, that might annihilate into gamma rays. Their predicted energy is in the accessible range of the MAGIC telescope. The expected gamma fluxes depend strongly on the density profiles in the innermost regions of the Dark Matter halos.

The prospects and strategies for indirect Dark Matter searches with the MAGIC Telescope are described. The observability and flux expectations from possible targets are discussed.

## 1. Introduction

The existence of Dark Matter is well established on scales from galaxies to the whole universe. Nevertheless, its nature is still unknown. Most of it cannot even be made of any of the known matter particles. A number of viable Weakly Interacting Massive Particle (WIMP) candidates have been proposed within different theoretical frameworks, mainly motivated by extensions of the standard model of particle physics (for a review see [1]). These include the widely studied models of supersymmetric (SUSY) Dark Matter [14]. Supersymmetric extensions of the standard model predict the existence of a good Dark Matter candidate, the neutralino $\chi$. In most models its mass is below a few TeV. Also models involving extra dimensions are discussed like Kaluza-Klein Dark Matter [12, 13].

Any WIMP candidate (SUSY or not) may be detected directly via elastic scattering off nuclei in a detector on Earth. There are several dedicated experiments already exploiting this detection technique, but they have not yet claimed any strong and solid detection (for a review see [2]). Complementary, WIMPs and especially SUSY neutralinos might annihilate in high-density Dark Matter environments and may be detected by their annihilation products. In particular, annihilation channels that produce gamma-rays are interesting because these are not deflected by magnetic fields and preserve the information of the original annihilation region, i.e. they act as tracers of the Dark Matter density distribution.

The expected mass range of SUSY neutralinos lies between about 50 GeV and a few TeV. Thus the continuum gamma-ray spectra from potential SUSY neutralino annihilation coincides well with the MAGIC energy region.

## 2. The MAGIC Experiment

The Major Atmospheric Imaging Cherenkov telescope (MAGIC [3]) is the largest Imaging Air Cherenkov Telescope (IACT). Located on the Canary Island La Palma at 2200m a.s.l, the telescope has a 17m diameter high reflectivity tessellated parabolic mirror dish, mounted on a light weight carbon fiber frame. It is equipped with a high efficiency 576-pixel photomultiplier camera, whose analogue signals are transported via optical fibers to the trigger electronics and the 300 MHz FADC readout. Its physics program comprises, among other topics, pulsars, supernova remnants, active galactic nuclei, micro-quasars, gamma-ray bursts and Dark Matter.

MAGIC has started observations in summer 2004, during the last phase of commissioning. Several known gamma sources were observed and analyzed like the Crab nebula, Mrk-421 and 1ES1959+650. A further challenge is the analysis of events below 100 GeV. The analysis methods are presently being adapted to these low energies. A second telescope, MAGIC-II, is being constructed and expected to be ready for data taking in the end of 2006. This will improve the angular and spectral resolution and flux sensitivity of the system.

## 3. Gamma-rays from neutralino annihilations

Neutralino annihilation can generate continuum $\gamma$-ray emission, via the process $\chi\chi \rightarrow q\bar{q}$. The subsequent decay of $\pi^0$-mesons created in the resulting quark jets produces a continuum of $\gamma$-rays. The expected annihilation $\gamma$-ray flux above an energy threshold $E_{\text{thresh}}$ arriving at Earth is given by:

$$\frac{dN_\gamma(E_\gamma > E_{\text{thresh}})}{dt\; dA\; d\Omega} = N_\gamma(E_\gamma > E_{\text{thresh}}) \cdot \frac{1}{2} \cdot \frac{\langle\sigma v\rangle}{4\pi m_\chi^2} \cdot \int_{\text{los}} \rho_\chi^2(\vec{r}(s,\Omega))ds \;, \tag{1}$$

where $\langle\sigma v\rangle$ is the thermally averaged annihilation cross section, $m_\chi$ the mass and $\rho_\chi$ the spatial density distribution of the hypothetical Dark Matter particles. $N_\gamma(E_\gamma > E_{\text{thresh}})$ is the gamma yield above the threshold energy per annihilation. The predicted flux depends on the SUSY parameters and on the spatial distribution of the Dark Matter. The energy spectrum of the produced gamma radiation has a very characteristic feature with a cut-off at the mass of the Dark Matter particle. Moreover, the flux should be absolutely stable in time.

As the expected flux is proportional to the Dark Matter density squared, high density Dark Matter regions are the most suitable places for indirect Dark Matter searches. Simulations and measurements of stellar dynamics indicate that the highest Dark Matter densities can be found in the central part of galaxies and Dark Matter dominated dwarf-spheroidal-satellite galaxies (with large mass-to-light ratio). Numerical simulations in a Cold Dark Matter framework predict a few universal DM halo profiles (for example see [4]). All of them differ mainly at low radii (pc scale), where simulation resolutions are at the very limit.

Combining the SUSY predictions with the models of the DM density profile for a specific object, the gamma flux from neutralino annihilations can be derived. The SUSY predictions are taken from a detailed scan of the parameter space assuming Minimal Supergravity (mSUGRA), a simple and widely studied scenario for supersymmetry breaking (for details see [5]). For a given choice of mSUGRA parameters the values of $m_\chi$, $\langle\sigma v\rangle$ and $N_\gamma$ are determined and the consistency with all observational constrains is checked. High DM density objects which are relatively nearby like the center of the Milky Way, its closest satellites and the nearby galaxies (M31,M87) are prime candidates for the indirect search for Dark Matter annihilation.

### 3.1 Galactic Center

The presence of a Dark Matter halo in the Milky Way Galaxy is well established by stellar dynamics [15]. In particular, stellar rotation curve data of the Milky Way can be fit with the universal DM profiles predicted by simulations [8, 7, 9]. In addition, the Dark Matter may be compressed due to the infall of baryons to the innermost region [5] of a galaxy creating a central spike of the Dark Matter density. This central Dark Matter spike would boost the expected gamma flux from neutralino annihilation in the center of the galaxy. Although this model of baryonic compression is based on observational data and in good agreement with cosmological simulations of the condensation of baryons [6], the existence of such a Dark Matter spike in the center of the Milky Way strongly depends on the Black Hole history during galaxy formation.

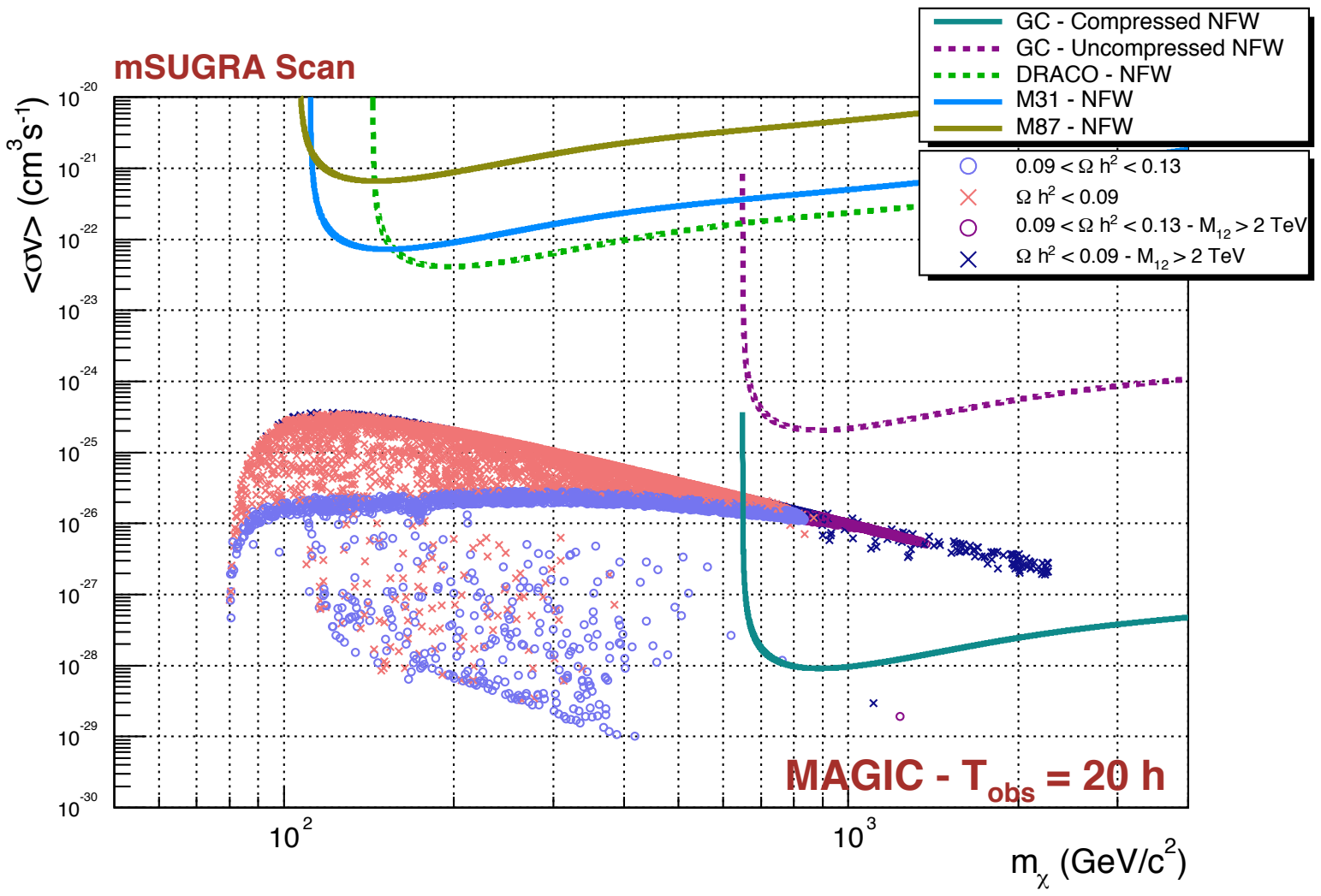

**Figure 1.** Expected exclusion limits for the four most promising sources of Dark Matter annihilation radiation for 20 hours of observation with MAGIC. The Galactic Center is expected to give the largest flux (lowest exclusion limits) amongst all sources.

For a comparison between the expected gamma fluxes from neutralino annihilation and the MAGIC sensitivity both the uncompressed NFW DM halo model [7] and the adiabatic contracted NFW profile [5] are considered.

### 3.2 Draco dwarf spheroidal and nearby galaxies

The Milky Way is surrounded by a number of small and faint companion galaxies. These dwarf satellites are by far the most Dark Matter dominated known objects, with Mass-to-Light ratios up to 300. Draco is the most DM dominated dwarf satellite. DM density profiles derived from Draco stars cannot differentiate between cusped or cored profiles in the innermost region, as data are not available at small radial distances. Moreover, observational data disfavors tidal disruption effects, which may affect dramatically the DM distribution in Draco. We adopt the recent cusped DM model which includes new Draco rotation data [8].

Moreover, we adopted NFW models for the nearby galaxy M31 [9] and the Virgo Cluster [10]. These profiles do not take into account any enhancement effect, like adiabatic contraction or presence of DM substructures.

## 4. Summary

Comparing the expected gamma ray flux from neutralino annihilation in the considered candidate sources with the MAGIC sensitivity [17], expected exclusion limits can be derived. Figure 1 shows expected exclusion limits for 20 hours of MAGIC observations in the mSUGRA plane $N_\gamma(E_\gamma > E_{\rm thresh})\langle\sigma v\rangle$ vs. $m_\chi$ for the four most promising sources considered. The nominal energy threshold $E_{\rm thresh}$ has been conservatively assumed to be 100 GeV. The change in energy threshold and effective collection area of the telescope with the zenith angle of the observation have been taken into account.

The expected fluxes are rather low and depend strongly on the innermost density region of the DM halos considered. The detection of a DM $\gamma$-ray signal from the Galactic Center is possible in case of a very high density DM halo, like the one predicted by adiabatic contraction processes. Improvements on the $E_{\text{thresh}}$ could allow to test a significant portion of the SUSY parameter space. The gamma flux from the Galactic Center as measured by the HESS experiment is far above the theoretical expectations and extends to energies above 10 TeV [16]. Thus only part of this flux may be due to the annihilation of SUSY-neutralino Dark Matter particles. Nevertheless, other models like Kaluza-Klein Dark Matter are not ruled out. It is interesting to investigate and characterize the observed gamma radiation to constrain the nature of the emission. Due to the large zenith angle for Galactic Center observations, MAGIC will have a large energy threshold but also a large collection area and good statistics at the highest energies. The Galactic Center was observed recently (May-July 2005) and data are being analyzed [11].

In the long term we consider Draco as a plausible candidate for Dark Matter inspired observations. Conservative scenarios give low fluxes which are not detectable by MAGIC in a reasonable observation time. However, there are several factors that might enhance the expected flux from neutralino annihilations in Draco. Other Dark Matter particles, like Kaluza-Klein particles, may produce higher gamma-rays fluxes. Draco is the most DM dominated dwarf (M/L up to 300) and an object where no other $\gamma$-ray emission is expected. Low zenith angle observations will preserve the nominal (low) $E_{\text{thresh}}$ of the MAGIC telescope. Moreover, there are no known high energy $\gamma$-ray sources in the FOV which could compete with the predicted gamma flux from Dark Matter annihilation.

### 4.1 Acknowledgments

The authors thank A. Moralejo for fruitful discussions about the MAGIC sensitivity.

# Tests of a Prototype Multiplexed Fiber-Optic Ultra-fast FADC Data Acquisition System for the MAGIC Telescope

H. Bartko, F. Goebel, R. Mirzoyan, W. Pimpl, M. Teshima
*Max-Planck-Institut für Physik, Föhringer Ring 6, 80805 Munich, Germany*
Presenter: H. Bartko (hbartko@mppmu.mpg.de), ger-bartko-H-abs1-og27

Ground-based Atmospheric Air Cherenkov Telescopes (ACTs) are successfully used to observe very high energy (VHE) gamma rays from celestial objects. The light of the night sky (LONS) is a strong background for these telescopes. The gamma ray pulses being very short, an ultra-fast read-out of an ACT can minimize the influence of the LONS. This allows one to lower the so-called tail cuts of the shower image and the analysis energy threshold. It could also help to suppress other unwanted backgrounds.
Fast 'flash' analog-to-digital converters (FADCs) with GSamples/s are available commercially; they are, however, very expensive and power consuming. Here we present a novel technique of Fiber-Optic Multiplexing which uses a single 2 GSamples/s FADC to digitize 16 read-out channels consecutively. The analog signals are delayed by using optical fibers. The multiplexed (MUX) FADC read-out reduces the cost by about 85% compared to using one ultra-fast FADC per read-out channel.
Two prototype multiplexers, each digitizing data from 16 channels, were built and tested. The new system will be implemented for the read-out of the 17 m diameter MAGIC telescope camera.

## 1. Introduction

MAGIC is the world-wide largest Imaging Air Cherenkov Telescope (IACT). It aims at studying gamma ray emission from the high energy phenomena and the violent physics processes in the universe, at the lowest energy threshold among existing IACTs [2]. The camera of the MAGIC Telescope consists of 576 Photomultiplier tubes (PMTs), which deliver via an analog-optical link about 2 ns FWHM fast pulses to the experimental control house [4]. The currently used read-out system [1] is relatively slow (300 MSamples/s). To record the pulse shape in detail, an artificial pulse stretching to about 6.5 ns FWHM is used. This causes more light of the night sky to be integrated, which acts as additional noise. Thus the analysis energy threshold of the telescope is limited, and the selection efficiency of the gamma signal from different backgrounds is reduced.

For the fast Cherenkov pulses (2 ns FWHM), a FADC with 2 GSamples/s can provide at least four sampling points. This permits a reasonable reconstruction of the pulse shape. Monte Carlo (MC) based simulations predict different time structures for gamma and hadron induced shower images as well as for images of single muons. The timing information is therefore expected to improve the separation of gamma events from the background events [3]. Such an ultra-fast read-out can improve the performance of MAGIC. The improved sensitivity and the lower analysis energy threshold will extend the observation range of MAGIC, and allow one to search for weak sources at high redshifts.

A few FADC products with $\geq$ 2 GSamples/s and a bandwidth $\geq$ 500 MHz are available commercially; they are, however, very expensive and power-consuming. To reduce the cost of an ultra-fast read-out system, a 2 GSamples/s read-out system has been developed at the Max-Planck-Institut für Physik in Munich. It uses the novel technique of Fiber-Optic Multiplexing [10], an approach possible because the signal duration (few ns) and the trigger frequency (typically $\sim$1 kHz) result in a very low duty cycle for the digitizer. The new technique uses a single FADC of 700 MHz bandwidth, 10 bit resolution and of 2 GSamples/s to digitize 16 read-out channels consecutively. The analog signals are delayed by using optical fibers. A trigger signal is generated using a fraction of the light, which is branched off by fiber-optic light splitters before the delay fibers. All optical components and the FADCs are commercially available, while the multiplexer electronics has been developed at the MPI in Munich.

## 2. The Ultra-fast Fiber-Optic MUX-FADC Data Acquisition System

The basic idea of the MUX-FADC system is to “pack” the signals of many channels into a single FADC channel. The block diagram of the MUX-FADC system is shown in figure 1. The ultrafast fiber-optic multiplexer consists of three main components: fiber-optic delays and splitters, multiplexer electronics (fast switches and controllers) and ultra-fast FADCs.

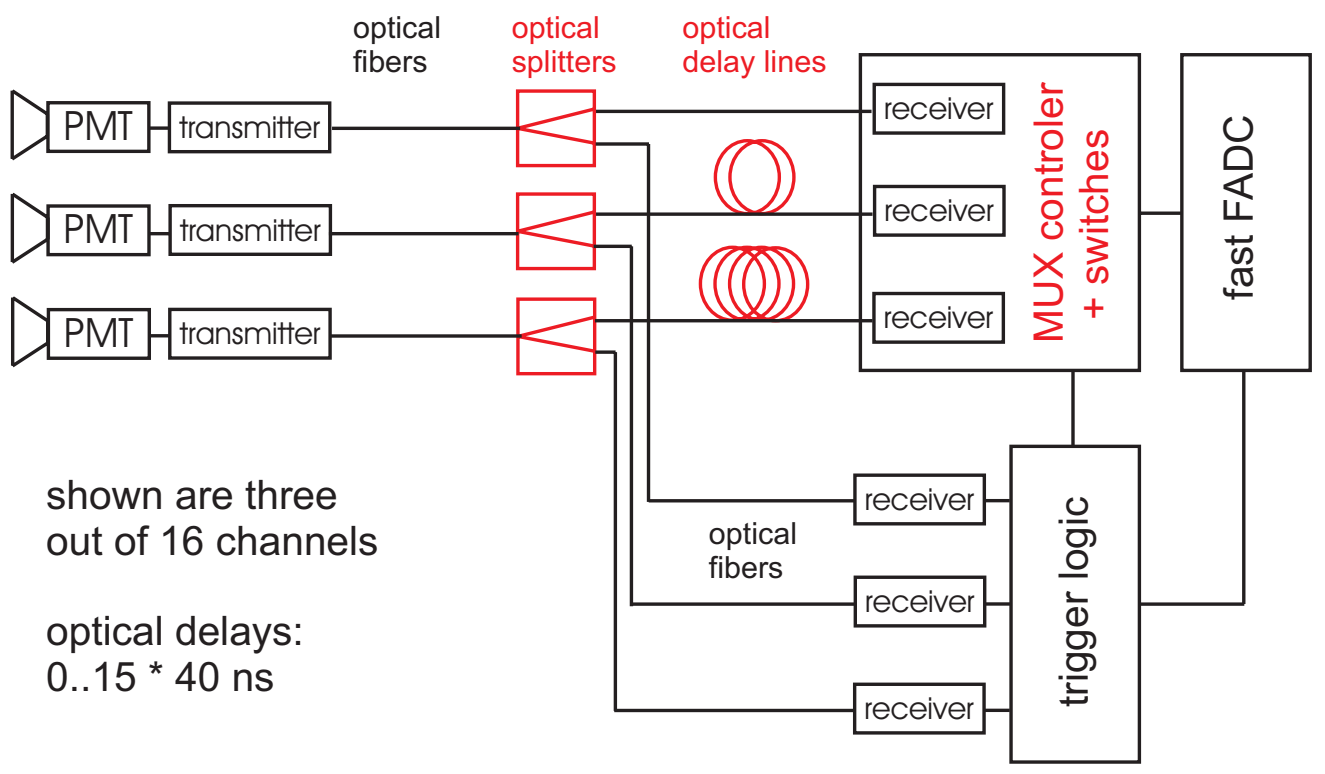

**Figure 1.** *Schematic diagram of the multiplexed fiber-optic ultra-fast FADC read-out. See text.*

After the analog optical link between the MAGIC PMT camera and the counting house the optical signals are split into two parts. One part of the split signal is used as an input to the trigger logic. The other part is used for FADC measurements after passing through a fiber-optic delay line of a channel-specific length.

The multiplexer electronics allows only the signal of one channel at a time to pass through and be digitized by the FADC. The other channels are attenuated by more than 60 dB for the fast MAGIC signals. In this way one “packs” signals from different channels in a time sequence which can be digitized by a single FADC channel.

Because of the finite rise and fall times of the gate signals for the switches and because of some pick-up noise from the switch one has to allow for some switching time between the digitization of two consecutive channels. The gating time for each channel was set to 40 ns, of which the first and last 5 ns are affected by the switching process. For the use in MAGIC a 16 $\rightarrow$ 1 multiplexing ratio was chosen. 16 channels are read out by a single ultra-fast FADC channel. The technological part of the fiber-optic multiplexer is described in detail in reference [10].

## 3. Prototype Test in the MAGIC Telescope on La Palma

Two prototype MUX-FADC read-out modules for 32 close packed channels were tested as a read-out of the MAGIC telescope during two weeks in August/September 2004. They were integrated into the MAGIC read-out system allowing the simultaneous data taking with the current 300 MSamples/s read-out and the MUX-FADC prototype read-out. In order to acquire only events where the shower image is located in the 32 MUX-FADC channels, only these channels were enabled in the MAGIC trigger system.

In figure 2a one can see the pulse shape in a single pixel for a typical cosmics event. By overlaying the recorded FADC samples of many events after adjusting to the same arrival time, the average reconstructed pulse shapes can be calculated. Figure 2b shows the comparison of the average reconstructed pulse shapes recorded with the current 300 MSamples/s MAGIC FADCs, including the 6ns pulse stretching, and with the MUX-FADCs. The average reconstructed pulse shape for cosmics events has a FWHM of about 6.3 ns for the current FADC system and a FWHM of about 3.2 ns for the MUX-FADC system.

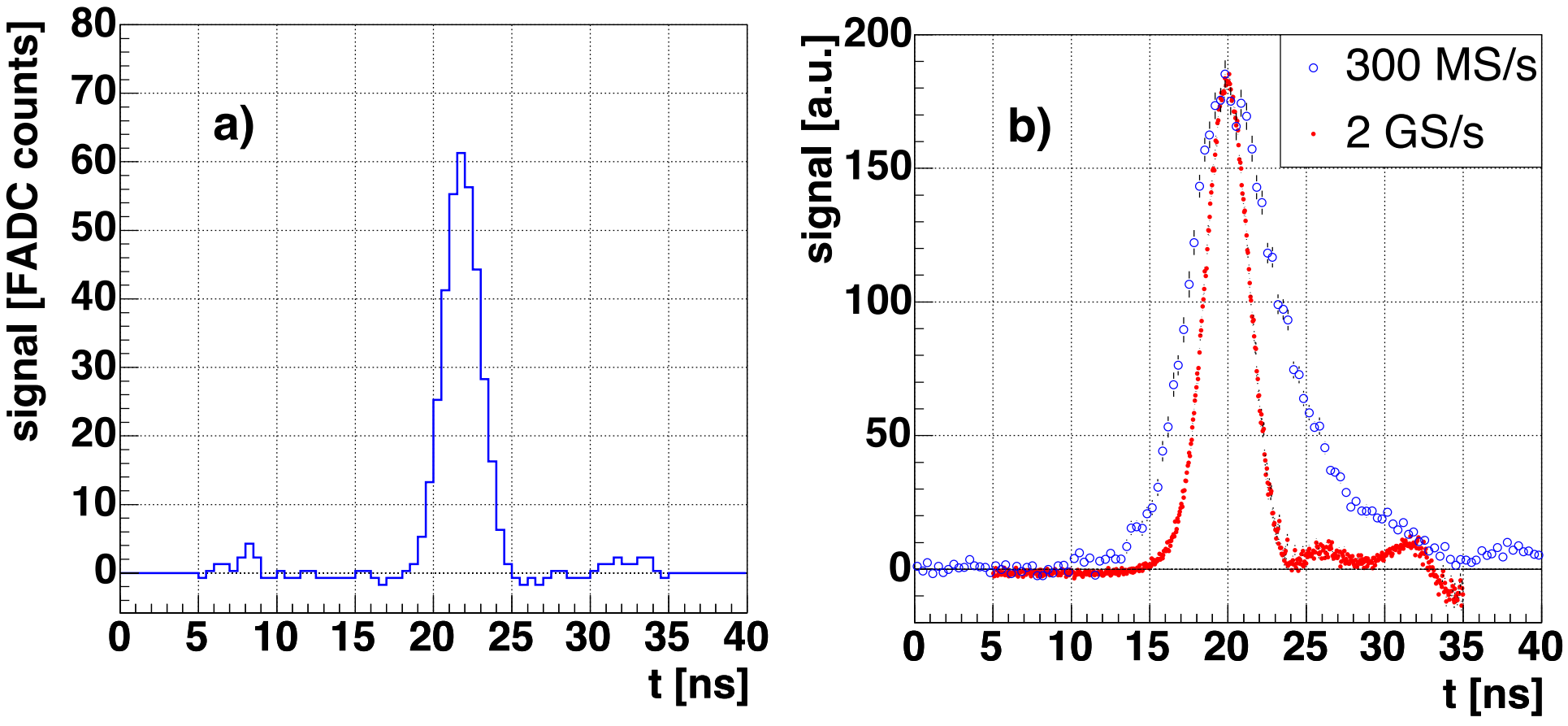

**Figure 2.** *a) Pulse shape in a single pixel for a typical cosmics event after pedestal subtraction. b) Comparison between the mean reconstructed pulse shapes recorded with the current MAGIC FADCs (open circles) and with the MUX-FADCs (full points).*

For the signal reconstruction a fixed number of FADC samples is integrated. The integration interval was chosen to be 4 FADC samples (corresponding to 4*3.33 ns = 13.33 ns) for the current MAGIC FADCs. For the MUX-FADCs a window size of 10 FADC samples is chosen, corresponding to a 5 ns integration window [7, 10]. The arrival time was calculated as the first moment of the FADC samples used for the charge integration.

The shorter integration time used for the pulse reconstruction with the MUX-FADC system yields a reduction of the effective integrated noise by about 40% for the MUX-FADC system compared to the current FADC system. Using the new MUX-FADC system the noise contributions due to the LONS may even be resolved into individual LONS pulses. Thereby a single photo electron spectum could be obtained and used for in-situ calibration.

The MAGIC camera can be isochronously and uniformely illuminated by intensity controlled fast LED light pulsers of different colors [8]. The event to event variation of the timing difference between two read-out channels for the LED pulser provides a measure of the timing accuracy. The timing accuracy strongly depends on the signal to noise ratio and the width of the input light pulse. The MUX-FADCs yield a better timing resolution (0.35 ns) by more than a factor of three compared to the current FADC (1.3 ns) system using the simple and stable timing extraction algorithm. Even better timing resolutions may be achieved with dedicated algorithms that fit the pulse shape [7].

## 4. Discussion

The ultra-fast fiber-optic multiplexed FADC prototype read-out system was successfully tested during normal observations of the MAGIC telescope in La Palma. The ultra-fast FADC read-out has grown to a mature technology which is ready for the use as a standard read-out system for the MAGIC telescope and other high-speed data acquisition applications.

The MUX-FADC read-out reduces the costs by about 85% compared to using one ultra-fast FADC per read-out channel. Also the power consumption of the read-out system is greatly reduced.

The ultra-fast MUX-FADC system allows to use a shorter integration window for the Cherenkov pulses. The reduction of the pulse integration window from 13.33 ns (4 samples with 3.33 ns per sample) for the current MAGIC FADC system to 5 ns (10 samples with 0.5 ns per sample) for the MUX-FADC system corresponds to a reduction of the integrated LONS charge by a factor of about 2.7. Consequently, the RMS noise of the LONS is reduced by about 40%.

A reduction in the noise RMS translates into lower image cleaning levels [10], a larger part of the shower image (a shower image of a higher signal to noise ratio) can be used to calculate Hillas parameters [9]. This is especially important for low energy events where the signals of only a few pixels are above the image cleaning levels. This will allow the reduction of the analysis energy threshold of the MAGIC telescope.

The ultra-fast FADC system also provides an improved resolution of the timing structure of the shower images. As indicated by MC simulations [3] gamma showers, cosmic ray showers and the so called single muon events have different timing structures. Thus the ultra-fast FADC read-out can enhance the separation power of gamma showers from backgrounds.

After the successful prototype test of the ultra-fast MUX-FADC read-out system it is ready to be installed as a future read-out of the MAGIC telescope.

## Acknowledgments

The authors thank R. Maier and T. Dettlaf from the electronics workshop of MPI. We also acknowledge the very good collaboration with the companies Acqiris and Sachsenkabel.

# Development of HPDs with an 18-mm-diameter GaAsP photo cathode for the MAGIC-II project

M.Hayashida[a], J.Hose[a], M.Laatiaoui[a], E.Lorenz[a], R.Mirzoyan[a], M.Teshima[a], A.Fukasawa[b], Y.Hotta[b], M.Errando[c], M.Martinez[c].

*(a) Max-Planck Institut für Physik, Föhringer Ring 6, 80805 München, Germany*

*(b) Hamamatsu Photonics, 314-5,Shimokanzo, 438-0193 Iwata, Japan*

*(c) Institut de Fisica d'Altes Energies, UAB, Barcelona, Spain*

Presenter: M. Hayashida (mahaya@mppmu.mpg.de), ger-hayashida-M-abs1-og27-oral

A new type of Hybrid PhotoDetectors (HPDs) with an 18-mm-diameter GaAsP photo cathode was developed in order to lower the energy threshold of the MAGIC-II project down to 15 GeV. The peak value of the Quantum Efficiency (QE) of these novel sensors reaches ~ 50 % at around 500 nm. Application of the wavelength shifting technique can further enhance the sensitivity in the UV region by another ~ 10 %. Compared to the currently used classical Photo Multiplier Tubes (PMT) with bialkali photo cathodes, the new HPDs provide practically twice as high photon conversion efficiency. Simulation studies of the new HPDs indicate that they have a sufficiently long lifetime for being used in the MAGIC telescope imaging camera. In this report we present the evaluation results and the performance of these new photo detectors.

## 1. Introduction

The development of the Imaging Atmospheric Cherenkov Telescopes (IACTs) technique has successfully established them as powerful tools for ground-based multi-GeV and TeV gamma-ray astronomy. Detection of gamma-ray fluxes at very low energies from 10 GeV to a few hundred GeV can be very interesting for the study of high red-shift objects such as active galactic nuclei and gamma-ray bursts. Because of the interaction of gammas from distant sources with the cosmic infrared background radiation fields, their measured intensity on the earth can be strongly attenuated. Only at very low energies is the universe becoming transparent for gammas. Use of photo sensors with higher photon conversion efficiency can be considered as an economic method to lower the threshold setting of telescopes.

The MAGIC (Major Atmospheric Gamma-Imaging Cherenkov) telescope [1], with a reflector diameter of 17 m, is the world's largest IACT. Since fall 2003 it has been in operation on the Canary Islands of La Palma (28.75° N, 17.90° W and 2200 m a.s.l.). In order to further lower the threshold setting our project will be upgraded by building the second 17-m diameter telescope (MAGIC-II) located at 85-m distance from the first telescope. In the MAGIC-II Project, one of the key tasks is to develop high QE Hybrid PhotoDetectors (HPDs) with a GaAsP photo cathode [2][3] as an alternative photo sensor to PMTs that are used in IACTs. For a high QE photo cathode, the Negative Electron Affinity (NEA) photo cathodes are regarded as the preferred candidates. Especially the NEA GaAsP type photo cathode is a prime candidate to be used in IACT photo sensors because of its high blue sensitivity.

An HPD consists of a photo cathode and of an Avalanche Diode (AD) serving as an anode. When applying a ~ 8kV high tension to the photo cathode, the photo electrons are accelerated in the high electric field and impinge onto the AD producing ~ 1600 electron-hole pairs. This is the so-called electron bombardment amplification. Those electrons subsequently induce avalanches in the active volume of AD and provide an additional gain of ~ 30-50 when a bias voltage of a few hundred volts is applied.

In the conventional HPDs, the size of the GaAsP photo cathode is too small (< 8mm) to be used as a pixel element in the MAGIC telescope camera (the necessary pixel size is 30 mm). Recently, together with Hamamatsu Photonics, we succeeded to produce HPDs with a GaAsP photo cathode of 18 mm size. By using non-imaging light concentrators like, for example, Winston cones, one can efficiently compress the light flux from the necessary 30 mm pixel input size to the 18 mm size of the above mentioned HPDs.

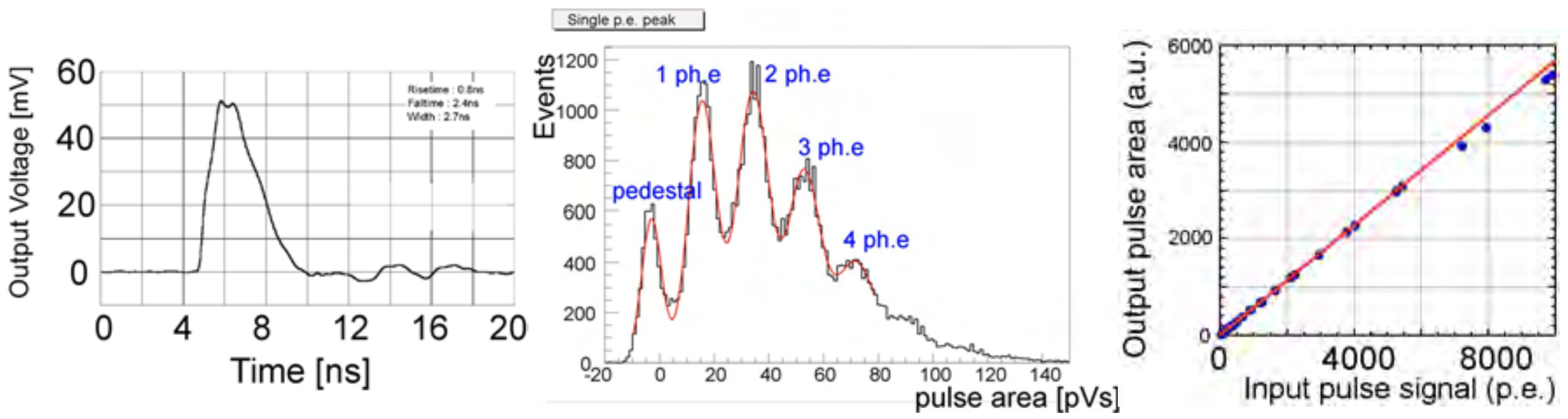

**Figure 1**. Examples of the measurement results.
**[(a) left]** The output signal for a very fast laser diode pulse. Photo cathode voltage; -8.0 kV, AD bias voltage; 330 V.
**[(b) middle]** Signal amplitude resolution. The peaks correspond to pedestal, 1 p.e., 2 p.e., 3 p.e. and 4 p.e..
**[(c) right]** Dynamic range measured by comparing the output signal area with input pulse signal of up to ~ 10,000 p.e..

## 2. Results and Discussions

HPDs including AD provide an overall gain of (3-8) x $10^4$. Due to the high signal gain of the first stage, HPDs have a very good amplitude resolution. Cherenkov light flashes from gamma-ray air showers have a time spread of 2-3 ns. Fast response is required in order to reduce the contribution from light of the night sky (LONS) for a short signal integration time. The QE of the HPD is dependent on the wavelength. Application of the Wavelength Shifting (WLS) technique can provide an increase in sensitivity in the UV region [4], where Cherenkov photons from air showers are more abundant. Although the photo sensors of IACTs are constantly exposed to the LONS during the operation, the photo cathodes of the light detectors shall not lose their sensitivity over several years of operation.

For the measurement of below, the laser diode (PDL 800B, PicoQuant GmbH) was used as the pulse generator. Its wavelength is 393 nm and the time width is several tens of ps (FWHM). A fast pre-amplifier (voltage gain 32.0 dB at 500 MHz) was used to amplify the HPD signal. The signal was acquired by a 1 Gsample/s digital oscilloscope (LC564A, LeCroy).

- **Gain:** the electron bombardment gain was measured at a photo cathode voltage of up to –8.5 kV. In the range of a few kV the gain is rising slowly due to the energy loss in the passive layer at the AD entrance window. Above 4 kV, the gain shows a linear relation with the photo cathode voltage. It reaches 650 at –5.0 kV and 1600 at –8.0 kV. On the other hand, the avalanche gain of the AD is 30 at 320 V. The breakdown voltage of the AD is about 350 V. Finally, the overall gain becomes about 50,000 at –8.0 kV of the photo cathode bias and 320 V for the AD bias voltage.
- **Time response:** Figure 1-(a) shows the output signal with –8.0 kV for the photo cathode and an AD bias voltage of 330 V. The intensity of the light was estimated to the hundreds of photoelectrons (p.e.). The output signal shows 2.7 ns FWHM. Although this can match the requirements for the IACT camera photo sensors, a faster response will nevertheless provide a shorter integration time.
- **Amplitude resolution:** As one can see in Figure 1-(b), multi-photoelectron peaks were well resolved at low light intensity. These peaks correspond to pedestal, 1 p.e., 2 p.e., 3 p.e. and 4 p.e. (from the left to right). The light level was adjusted to provide <1.95> p.e.. The relative amplitude resolution of the single p.e. distribution can be described by $\sigma_{gauss} \sim$ 18 % after removing the contamination due to the pedestal fluctuations.
- **Dynamic range:** In the small signal range, the multi-photoelectron peaks (in Figure 1-(b)) appear at regular interval within a 1% error. Figure 1-(c) shows the result of the dynamic range measured by the output signal area with input pulse signal of up to ~ 10,000 p.e.. The output signal area keeps a linear relation to the input pulse signal up to 5,000 p.e. and begins to deviate by 5% at 7,000 p.e..

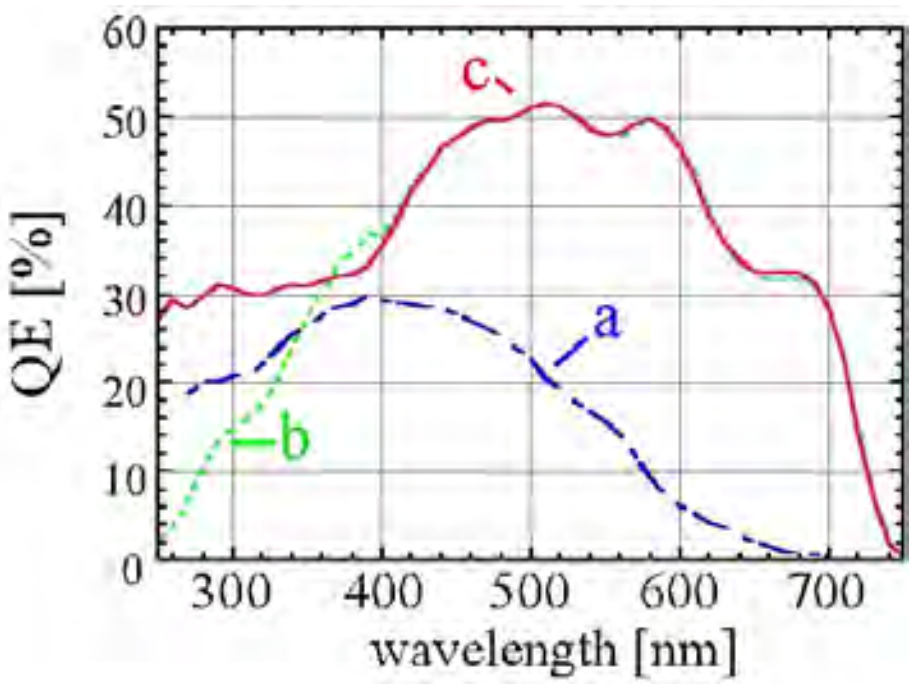

**Figure 2.** Measured QE spectra. **[broken (a)]** PMT ET-9116A with milky lacquer used in MAGIC-I [5]; **[dotted (b)]** non-coated HPD; **[solid (c)]** HPD with WLS

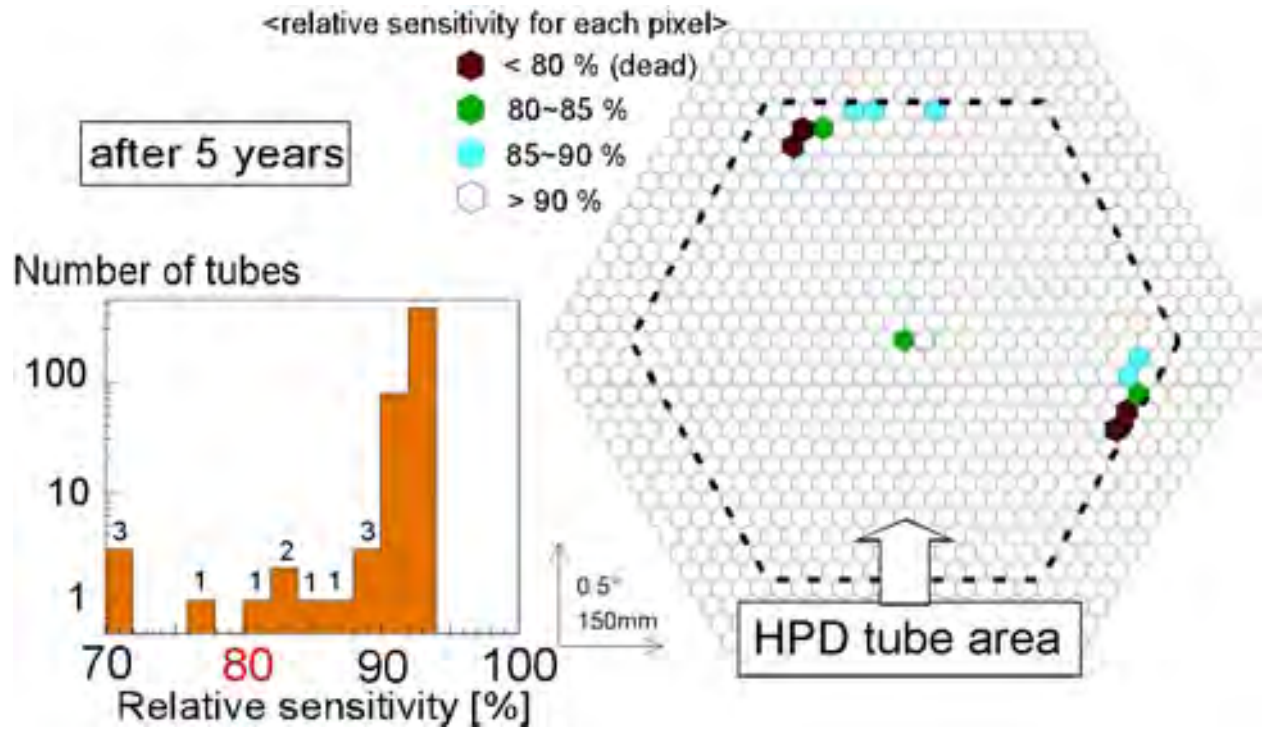

**Figure 3.** An example of a lifetime simulation study of relative photo cathode sensitivity compared with its initial value. This case shows the result after 5 years of operation.
**[(a)left]** The histogram of the number of tubes. **[(b)right]** Distribution of damaged pixels on the camera. The HPD tube area was assumed inside the dotted line. Each color represents a different relative sensitivity value. The damaged tubes mainly follow the track of ζ-Tauri.

| Zenith Angle | 0° | 25° | 45° | 60° |
|---|---|---|---|---|
| Non-coated | 1.90 | 1.92 | 2.00 | 2.14 |
| with WLS | 1.99 | 2.00 | 2.07 | 2.17 |

**Table 1.** Relative improvement of overall Cherenkov photon conversion efficiency when using GaAsP HPD compared to PMT.

**Quantum efficiency**

We measured the QE in the range from 250 to 750 nm in 10 nm intervals using a spectrometer. The current of the HPD was measured by shorting the AD anode with the cathode, and by applying –800 V to the photo cathode. A calibrated pin-photo diode (S6337-01, Hamamatsu Photonics, calibration accuracy 2 %) was used to get the absolute value of light intensity. Figure 2 shows the measured QE of the HPD photo cathode as well as the QE of the PMT used in the telescope of MAGIC-I [5] as a function of wavelength. The peak value reaches 51 % at around 500 nm.

The first tests of WLS technique were made with a mixture of 0.03 g POPOP, 0.03 g Butyl-PBD and 1.5 g Paraloid B72 dissolved in 20 ml of Toluene. This solution was dripped on the entrance window of the HPD, and thus obtaining thin and transparent layer. In Figure 2 the obtained QE spectra with and without application of the WLS are shown. The enhancement can be seen clearly below 360 nm. However, a small drop in sensitivity exists at around 400 nm because of the absorption by the shifter film. Further studies could provide better results.

In order to quantify the anticipated improvement when using HPDs for the MAGIC telescope, overall Cherenkov photon conversion efficiency was estimated by folding the QE and the expected Cherenkov photon spectrum from gamma-ray showers. Table 1 shows the improvement of the efficiency for four values of the observation zenith angle. The obtained value is normalized to that of the currently used PMT. The spectral peak position of the Cherenkov spectrum shifts towards the longer wavelengths at higher zenith angles, because the shorter wavelengths are stronger absorbed and scattered by the atmosphere. This calculation includes the differences in the collection efficiency of the different light guides (94 % for PMT, 87 % for HPD) and in the first anode (dynode) collection efficiency (90 % for PMT, 100 % for HPD). The results show that the total light conversion efficiency could be improved by about a factor 2 compared to the PMTs. At the high zenith angle range, the improvement could be even higher due to the red-extended sensitivity of the HPD. Depending on the observation zenith angle the WLS can provide an additional improvement of 3-9 %.

**Lifetime of the GaAsP photo cathode**

A GaAsP photo cathode lifetime was predicted empirically from the information of conventional image intensifiers with GaAsP photo cathodes. The lifetime is defined as the period after which the sensitivity degrades by 20 % from the initial value. On can assume that the lifetime only depends on the amount of total charge produced in the photo cathode. Then the lifetime corresponds to 3.5 mC of total charge from the photo cathode and about 100 C in the AD output charge with a gain of 30,000. The rate of p.e. due to LONS is estimated to be 0.45 p.e./ns/pixel (calculation based on the measured LONS spectrum [6]). At this LONS level the lifetime is ~13,500 hours.

In order to confirm the durability of the GaAsP photo cathode under real conditions, simulation of the starlight and LONS were performed. Figure 3-(b) represents an example of the camera pixel arrangement. In this simulation, we selected 10 known TeV sources (Crab Nebula, 3C66A, Mkn421, 1H1426, Mkn501, 1ES1959, BL-Lac, 1ES2344, Galactic Center, CasA) and picked up those of neighboring stars that are brighter than 11th magnitude in the V-band. The brightest star of all objects is ζ-Tauri (3.02 mag., located 1.13 deg. off the Crab). Observation time was assumed to be 100 hours per year for each of the sources, thus in total 1000 hours per year. Figure 3 shows the result after 5 years of operation. The relative photo cathode sensitivity compared with the initial value is shown. Figure 3-(a) is a histogram of the number of tubes as a function of the relative sensitivity. Most of the tubes still have a sensitivity of more than 90 %. The distribution of damaged pixels on the camera is shown in Figure 3-(b). The value of the relative sensitivity is shown by using the scale of four colors. The damage is mainly due to the intense light from ζ-Tauri. In the real operation, many other objects will be observed. But the expected number of the bright stars (brighter than 3.5 mag.) in the field of view of the HPD camera (~ 5 $deg^2$) is 0.034. Therefore the result of this simulation indicates that the HPD camera can keep good quality over several years with a small number of replacements of dead tubes.

## 3. Summary

The new type of HPD with the 18 mm GaAsP photo cathode is almost ready to be used in low threshold setting IACTs. The QE peak value reaches over 50 % at around 500 nm and the first tests of WLS technique demonstrated an increase of sensitivity in the UV region. Compared to the currently used PMTs in MAGIC-I, the overall Cherenkov photon conversion efficiency with the GaAsP photo cathode HPD is improved by a factor 2. This can be seen as equivalent of increasing the mirror diameter from 17 m to 24 m. Lifetime simulations showed that the GaAsP photo cathode is expected to have sufficiently a long lifetime to survive starlight and the light of night sky. In the next step further improvements of the QE as well as the development of a new AD with even higher gain and faster time response is planned. The GaAsP photo cathode lifetime studies are planned with a few tens of HPD tubes under different conditions in this summer. A HPD camera shall be built in 2006 in order to access the 15-GeV energy threshold for the MAGIC-II project.

## Acknowledgements

The MAGIC project is supported by the MPG (Max-Planck-Society in Germany), and the BMBF (Federal Ministry of Education and Research in Germany), the INFN (Italy), the CICYT (Spain) and the IAC (Instituto de Astrophysica de Canarias).

# Development and first results of the MAGIC central pixel system for optical observations.

F. Lucarelli[a,d], P. Antoranz[a], M. Asensio[a,c], J.A. Barrio[a], M. Camara[a], J.L. Contreras[a], R. de los Reyes[a], M.V. Fonseca[a], M. Lopez[a], J.M. Miranda[b], I. Oya[a] for the MAGIC collaboration.

*(a) Dpto. Física Atomica, Facultad de Ciencias Físicas, Universidad Complutense, 28040 Madrid, Spain.*
*(b) Dpto. Física Aplicada III, Facultad de Ciencias Físicas, Universidad Complutense, 28040 Madrid, Spain.*
*(c) Dept. Infra., I. Sistemas Aeroespaciales y Aerop., Universidad Politcnica, 28040 Madrid, Spain.*
*(d) Now at the Dip. di Fisica, Università degli Studi di Roma "La Sapienza", Ple. Aldo Moro 5, 00185 Roma, Italy.*
Presenter: F. Lucarelli (lucarel@gae.ucm.es), spa-lucarelli-F-abs1-og27-poster

The MAGIC telescope has been designed for the observation of the Čerenkov light generated in Extensive Air Showers. However, its 17 m. diameter and optical design makes it suitable for optical observations as well. In this work, we report on the development of a system based on the use of a dedicated photo-multiplier (PMT) for optical observations installed at the center of the MAGIC camera (the *central pixel*). An electro-optical system has been developed in order to transmit through optical fiber the PMT output signal to the counting room, where it is digitized and stored for off-line analysis. First tests of this system using the Crab nebula as calibration source show its optical pulsation.

## 1. Introduction

The MAGIC telescope [1] is an innovative detector aimed to detect very high- energy $\gamma$-rays from astrophysical sources using the Imaging Atmospheric Čerenkov technique. The telescope collects the very short flashes of atmospheric Čerenkov radiation (5-20 nsec in duration) emitted during the development of the Extended Atmospheric Showers (EAS) produced in the interaction of the cosmic $\gamma$-rays with the atmospheric nuclei.

The main characteristics of the telescope are its 17m. tessellated mirror, a system of analog signal transmission based on optical fiber and signal digital sampling with 300 MHz Flash ADCs. The telescope is especially designed to reach the extremely low energy threshold for primary $\gamma$-rays of 30 GeV. The main detector, or camera, is located at the focus of the telescope and consists of a matrix of 576 fast-response PMTs.

Besides the main $\gamma$-ray observations, the large collection area of the MAGIC telescope can also be used to perform optical observations of varying astronomical objects. That can be done by integrating the slow DC current output of a PMT. This technique was already applied by other Čerenkov telescopes (HEGRA CT1 [2], Whipple [3], Celeste [4], HESS [5]) which detected the optical pulsed emission from the Crab pulsar and estimated the photon content of the nebula surrounding the pulsar itself [2].

At commissioning time, the center of the MAGIC camera was deliberately left empty in order to host a dedicated PMT for optical observations. Such modified PMT, the so-called *central pixel*, was installed at the end of March '05 and tested successfully with the detection of the optical pulsed emission from the Crab pulsar.

In what follows, we will describe the main technical aspects of the installation of the central pixel and the tests performed with the observation of the Crab pulsar.

## 2. The central pixel

The PMT installed at the center of the MAGIC camera is a standard 1" MAGIC PMT, ET9116 [6], especially designed for fast pulsed-light detection. The DC branch of the pre-amplifier placed at the PMT base has been modified in order to have an integration constant $\tau = 0.5$msec. The overall tension was set to 1.08kV, corresponding to a gain of around 20k.

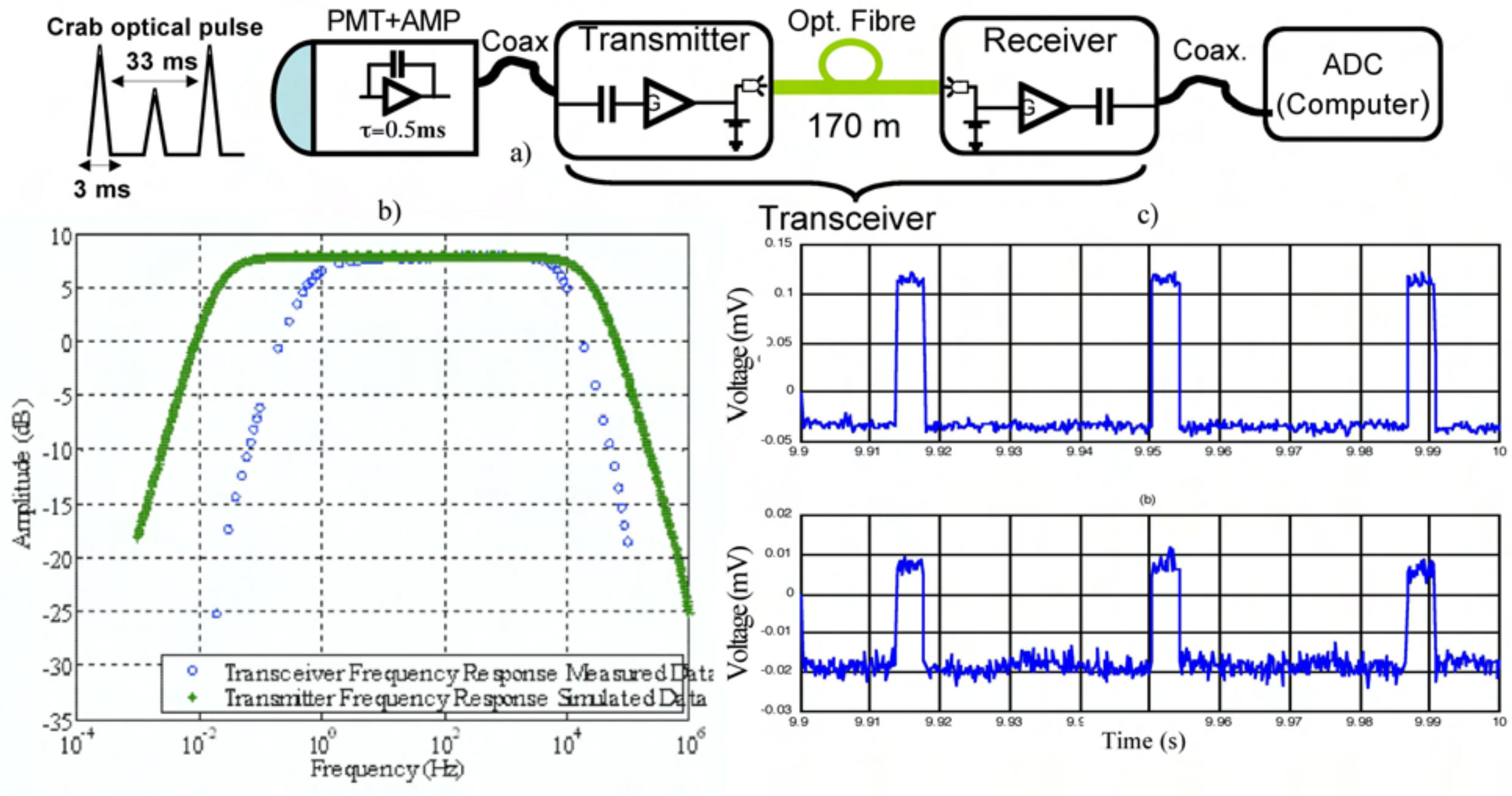

**Figure 1.** (a) Sketch of the transmission system for the central pixel. (b) Transceiver bandwidth. (c) On-site measurements at La Palma: pulsed signals as received at the end of the transceiver chain in two cases: for a 100 mV input signal at the transmitter (upper plot) and a 10mV input signal (lower plot).

The transmission of the DC output signal from the PMT base to the counting house for its digitalization and storing is made through optical fiber. The electro-optical transceiver (see Fig.1.a) was designed in order to transmit Crab-like signals, that is, analog pulsed signals with 10-100 Hz periods and widths of the order of milliseconds. Thus, instead of the high-speed VCSELs of the MAGIC DAQ chain (which detect the fast Čerenkov pulses), a wide dynamic range LED was used (Honeywell HFE 4050-014 [8]). The optical fiber (a graded index multimode fiber 50/125mm core/cladding, ($\lambda$=850nm.)) and the connectors (Diamond MAT E-2000) were already installed in the camera. The length of the optical fiber from the camera to the counting house is of about 170m.

At the emitter, the LED is set to an operation point of 40mA, thus providing linear operation over a wide dynamic range (up to 200 mV). The transceiver built allows to measure DC variations at the level of 0.2%.

At the receiver (inside the counting house), a pin diode with an analog pre-amplifier (Honeywell HFD3038-002 [9]) implements the optical-electric conversion. The bandwidth of the transceiver (Figure 1.b) is set from 1Hz to 4kHz, thus allowing transmission of Crab-like signals and rejecting high and low frequency noise (and DC background).

Figure 1.c shows the on-site tests of the whole transceiver system, where a series of pulsed signals (P=32Hz) were transmitted from the MAGIC camera to the counting house by feeding them directly into the transmitter by means of a pulse generator.

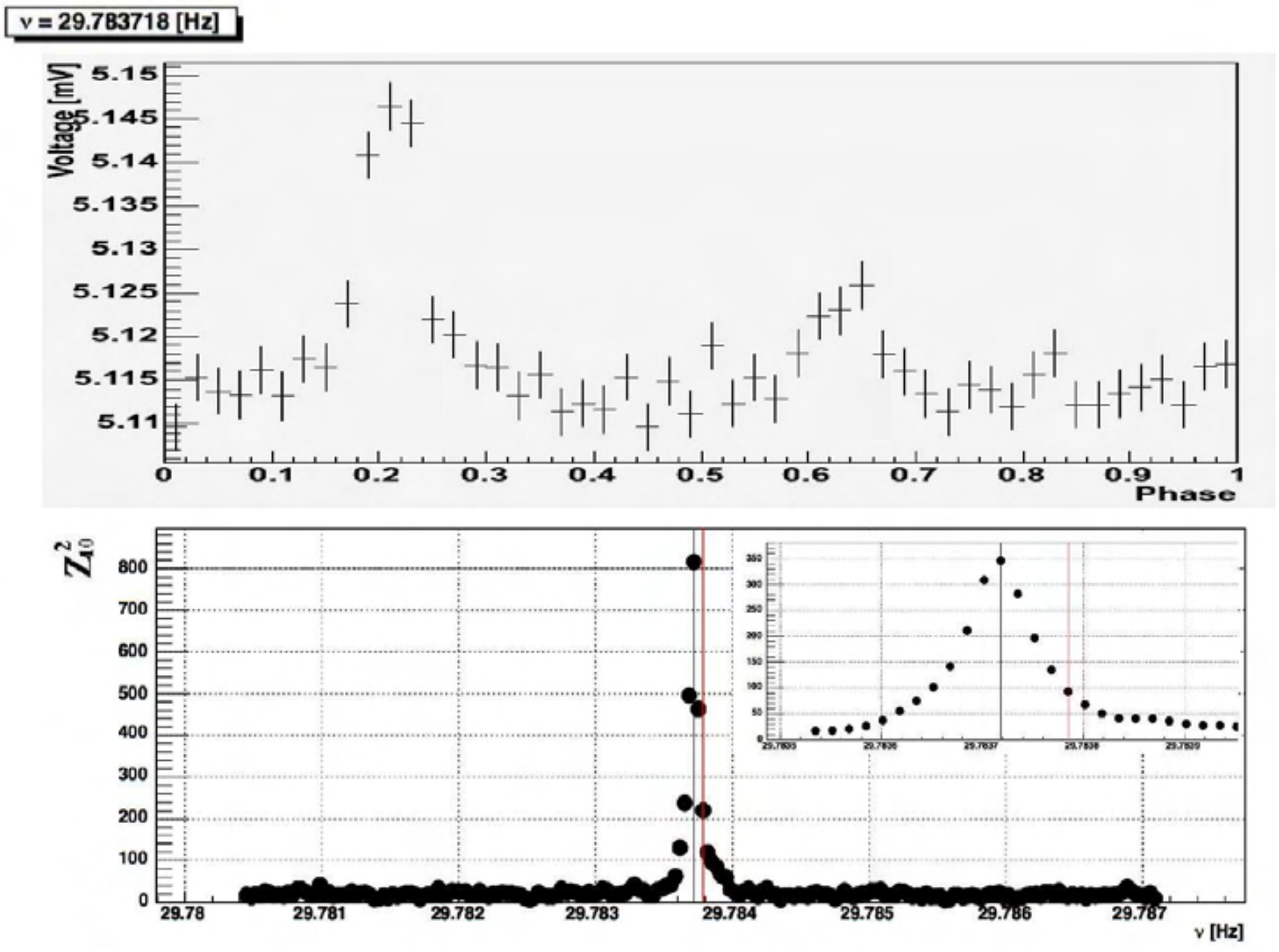

**Figure 2.** *Upper plot*: Crab light-curve. *Lower plot*: $Z^2_m$ statistical test. In the box on the left, a zoom of the region around the expected frequency.

## 3. Crab observations

The whole system has been installed at the end of March '05 and tested by observing the Crab pulsar. The DC current from the central pixel, converted to voltage, was digitized at a sampling frequency rate of 2 kHz by means of a 16-bits National Instruments PCI-6034E ADC card [10] and stored for off-line analysis. The ADC card was set to a resolution of 0.16$\mu$-volts.

A time stamp reflecting this frequency was associated at each event (the absolute event times were not available at that time) and transformed to an inertial reference frame, which is assumed as the solar barycenter system. For this transformation, the TEMPO software [12] has been used.

As the Crab pulses are completely embedded in noise (generated by the night sky background and the electronics), in order to observe its optical pulsation at the expected frequency, the corrected times must be folded with the radio ephemerides corresponding to the observation time provided by the Jodrell Bank observatory [13]. A phaseogram was produced around the expected Crab rotational frequency for each independent test frequency defined by the Independent Fourier Spacing IFS=1/$T_{obs}$. The folded intensities were tested against a uniform distribution by performing the $Z^2_m$ statistical test. Figure 2 (lower plot) shows the $Z^2_{10}$ test scanning different frequencies around the expected Crab frequency. A clear peak appeared at the frequency expected for the observation epoch. Figure 2 (upper plot) shows the well-known double-peaked lightcurve calculated at the maximum $Z^2_{10}$ value and after 20 min. of observation.

## 4. Conclusions

In this work, we have reported about the installation of the central pixel system of the MAGIC telescope, dedicated to optical observations. The central pixel has been tested successfully by the detection of the Crab optical pulsations. The minimum time requested for a $5\sigma$ detection was lower than 1 minute. The expected detection time was 30 sec [2]. However, due to extreme weather conditions preceding the weeks immediately before the observations, the optical conditions of the telescope were not the optimal. The pointing error of the telescope was estimated offline in about $0.1°$, while the Point Spread Function (PSF) was around $0.1°$ (FWHM). This limited the amount of collected light over the Central Pixel to about 7-9% of the total light emitted by the pulsar. Besides that, the observation was done at very high zenith angle, with a strong background due to zodiacal light. Recent alignments of the mirror have reduced the PSF to less than $0.05°$. Thus, we expect to improve the detection time by a factor at least 10.

The applications of the central pixel will be focused mainly in the simultaneous observations of the Crab pulsar in the optical and $\gamma$ regimes, in order to have real-time ephemerides for periodicity search in $\gamma$-rays. Before that, the central pixel will be also used to test the whole timing system of the MAGIC telescope [11]. Optical observations of flaring AGNs (Mrk 421, 501, ..) and X-ray binary systems (AE Aquarii) are also contemplated.

## 5. Acknowledgements

The authors wish to thank the financial support given by the CICYT (project FPA2003-9543-C02-01) to make this work and the IAC for providing excellent working conditions in La Palma. The MAGIC telescope is mainly supported by BMBF (Germany), CICYT (Spain), INFN and MURST (Italy).

# Concept of a Global Network of Cherenkov Telescopes and first joint observations with H.E.S.S. and MAGIC

D. Mazin[a], F. Goebel[a], D. Horns[b], G. Rowell[c], R.M. Wagner[a], S. Wagner[d]
for the MAGIC [e] and H.E.S.S. [f] Collaborations

*(a) Max-Planck-Institut für Physik, Föhringer Ring 6, D-80805 München, Germany*
*(b) Institut für Astronomie und Astrophysik, Universität Tübingen, D-72076, Germany*
*(c) Max-Planck-Institut für Kernphysik, D-69117 Heidelberg, Germany*
*(d) Landessternwarte Heidelberg, D-69117 Heidelberg, Germany*
*(e) http://wwwmagic.mppmu.mpg.de/ , (f) http://www.mpi-hd.mpg.de/hess/*

Presenter: D. Mazin (mazin@mppmu.mpg.de), ger-mazin-D-abs3-og23-poster

New generation Cherenkov telescopes cover a wide range of longitudes (137°E to 110°W) allowing continuous observations to follow transient sources. Given the close match in longitude of the MAGIC (17.9°W) and H.E.S.S. (16°E) sites, simultaneous observations at greatly differing zenith angles are also feasible. The measurable energy range can thus be extended beyond what is accessible to individual instruments. The planning and coordination of world-wide observations is challenging and requires close interaction between the different collaborations. The potential of Global Network of Cherenkov Telescopes (GNCT) campaigns for blazar physics and studies of the energy dependent absorption of very high energy (VHE) $\gamma$-rays on the extragalactic background light are discussed. Also, first results from a joint H.E.S.S. and MAGIC observation of Mkn 421 in 2004 are presented.

## 1. Introduction

The new generation of Imaging Air Cherenkov telescopes offers a unique opportunity to perform joint observation campaigns. Due to their similar longitude the fully operational H.E.S.S. and MAGIC telescopes allow simultaneous observations, which will greatly improve their capability to study variable VHE $\gamma$-ray sources and extend the energy range covered.

The High Energy Stereoscopic System (H.E.S.S. ) is located in Namibia (23°16′S, 16°30′E). It consists of 4 identical 13 m diameter Cherenkov telescopes with 107 $m^2$ tessellated and automatically adjustable glass mirror facets [1]. The energy threshold achieved for observations close to the zenith is around 100 GeV with an angular resolution of better than 0.1° for individual events. For large zenith angles the threshold energy increases to 1.2 TeV at 60° zenith.

The 17 m diameter MAGIC telescope is located on the Canary Island of La Palma (28°30′N, 17°53′W) at an altitude of 2200 m a.s.l. MAGIC is currently a stand-alone instrument with a second telescope under construction. Owing to its novel technologies and large mirror area, and its fine granulated camera with high quantum efficiency PMTs, MAGIC was designed to detect VHE $\gamma$-rays of energies down to 30 GeV [2]. Its sensitivity permits the detection of signals from Crab-like sources within a few minutes. MAGIC started regular observations in August 2004.

## 2. Observational prospects

The combination of similar longitude but widely different latitudes of the H.E.S.S. and MAGIC sites allows one to carry out simultaneously small and large or very large zenith angle observations for sources with dec-

.

lination $|\delta| > 20°$ like e.g. Mkn 421 ($\delta = +38°12'32''$), Mkn 501 ($\delta = +39°45'37''$), PKS 2155-304 ($\delta = -30°13'32''$), PKS 2005-489 ($\delta = -48°49'54''$), and H1426+428 ($\delta = 42°40'21''$). Observations at low zenith angles allow measurements with a low energy threshold. For large zenith angles the Cherenkov light cone illuminates a large region on the ground. Due to the reduced photon density this results in a higher energy threshold. On the other hand the effective collection area is significantly increased which improves the sensitivity for $\gamma$-rays at high energies, where the fluxes are very low. Figure 1 illustrates the observational situation between two facilities located as MAGIC and H.E.S.S.

Observatories at similar latitude but different longitude (e.g. MAGIC in La Palma and VERITAS in Arizona or H.E.S.S. in Namibia and CANGAROO III in Australia) allow a different observation strategy. Follow-up observations of variable sources effectively allow one to overcome the long gaps in observation time of ground based Cherenkov detectors during day time.

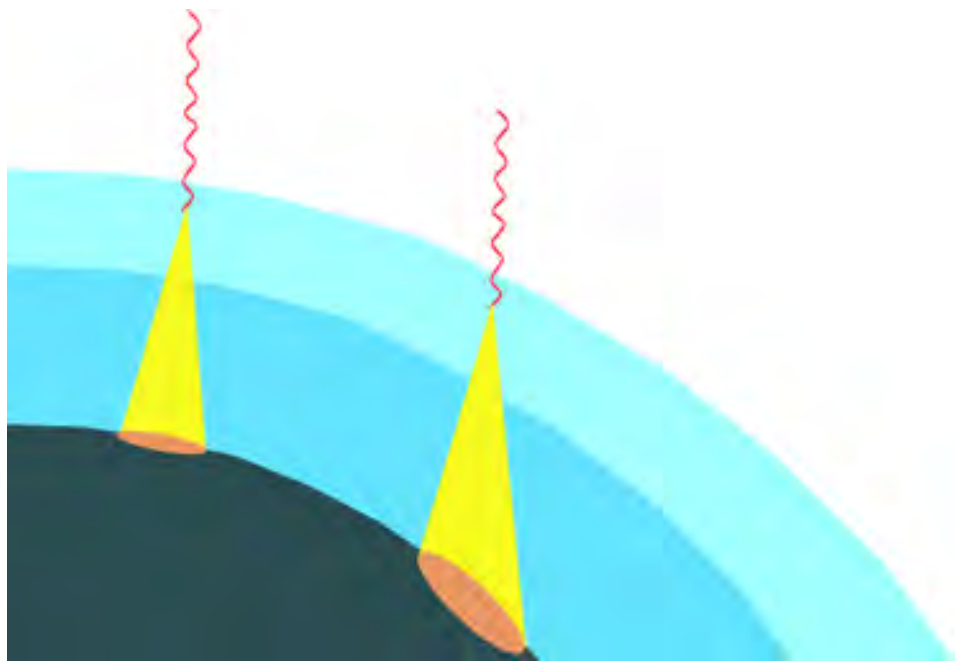

**Figure 1.** For two observatories at different latitudes, the showers will be observed under different inclination angles. Observations at high inclination angles result in a large collection area but also a high energy threshold.

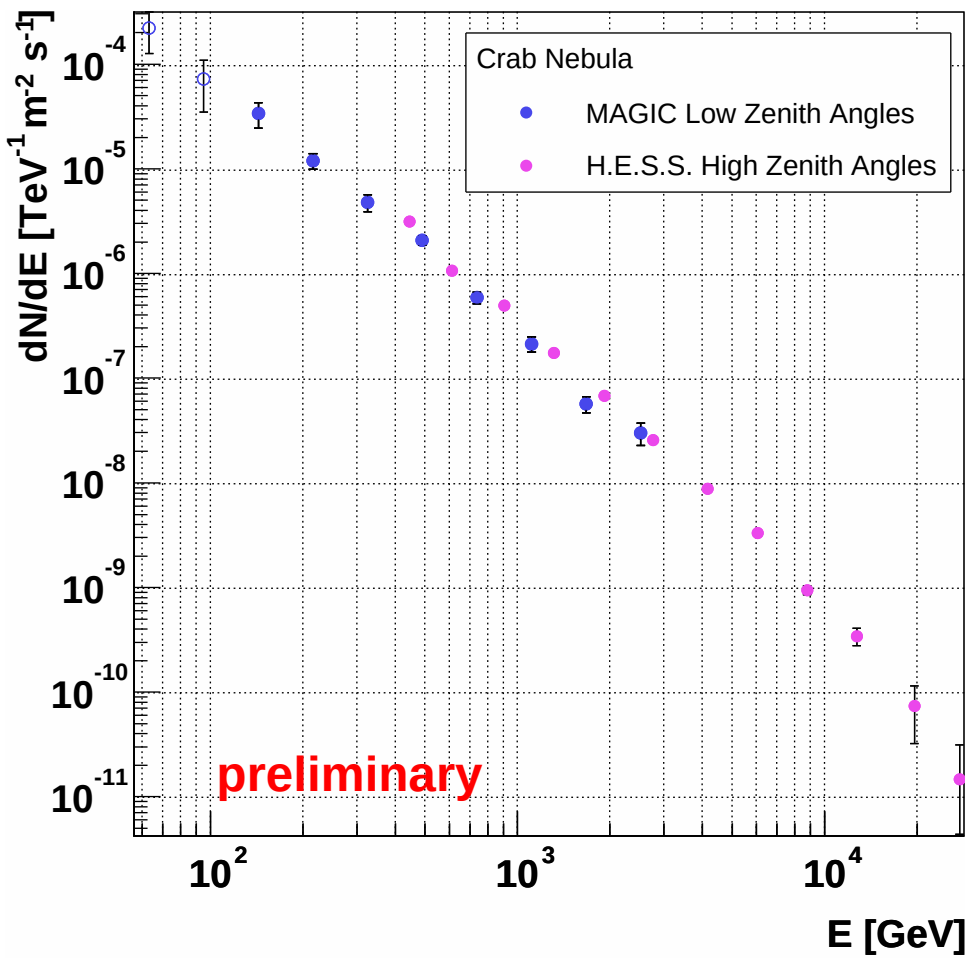

**Figure 2.** Combined energy spectrum from the Crab nebula taken around culmination. The large overlap of the energy spectra allows a good cross-calibration.

## 3. Physics goals

### GeV-TeV spectra cover 3 decades in energy

For simultaneous observations of bright sources ($F \geq F_{\rm Crab}$), it is possible to obtain a differential energy spectrum between 30 GeV and 30 TeV within a few hours of observation. In Figure 2, an energy spectrum of the Crab Nebula combining MAGIC and H.E.S.S. data is shown. The H.E.S.S. results have been obtained using 10 h of data from 2003/2004 at mean zenith angle of 45°. For the MAGIC measurements 2 hours of data taken in 2004 at a mean zenith angle of 15° have been used. The Crab Nebula is a standard candle for GeV-TeV emitting a constant flux of $\gamma$-rays. Therefore, data taken at different times can be compared. Since the Crab Nebula has a declination of 22°, the energy thresholds of the two telescopes, while differing by a factor of 7, still allow considerable overlap in the energy spectra. It is therefore an ideal source to cross calibrate the two instruments in the overlapping energy range. Figure 2 shows good agreement and indicates small systematic uncertainties of the independent detector calibrations.

**TeV blazars – catching rapid broadband variability**

TeV blazars are known to exhibit strong variations in their $\gamma$-ray emission. The intensity may change by more than one order of magnitude and large variations on time scales down to 10 minutes have been measured. Also the shape of the energy spectrum has been observed to change with the integral flux (see Figure 3 *a* for an example of the expected variation of the spectral energy distribution). A precise measurement of the energy spectra during different states of the source is essential to understand the physics of the highly relativistic plasma and its interaction with the ambient medium. Simultaneous measurements covering three energy decades as proposed here are an ideal tool to study the rapid broadband variabilities of TeV blazars. We expect a strong improvement in the understanding of the acceleration and cooling processes during flares of these objects. Good candidates for these studies are Mkn 421 ($z = 0.031$) and Mkn 501 ($z = 0.034$). They show strong flares reaching flux levels several times higher than that of the Crab Nebula.

**Measuring the optical depth due to pair creation processes**

$\gamma$-rays emitted from far distant galaxies are absorbed via pair-production by interaction with photons of the optical / near IR extragalactic background light (EBL): $\gamma_{\rm TeV}\gamma_{\rm EBL} \rightarrow e^+e^-$.
The expected density of the EBL is a matter of debate [3, 4, 5] and difficult to measure directly [6]. The measured spectra of distant GeV-TeV sources can be used to determine (or at least to constrain) the energy density profile of the EBL. The task is rather difficult since it requires knowledge of the intrinsic spectra of the GeV-TeV $\gamma$-ray emitters. However, as shown in Figure 3 *b* for a specific red shift range, the expected optical depth ($\tau$) is very small ($\tau < 1$) up to a few 100 GeV and then increases dramatically ($\tau > 1$) beyond TeV energies. Observations covering a broad band in energy would allow simultaneous measurement of the intrinsic spectrum of the source at low energies and at high energies where the apparent spectrum is modified by absorption due to pair production processes. The best known candidate for joint observations so far is H1426+428 ($z = 0.129$) with an intrinsically hard spectrum, so that the TeV component can still be measured. We expect to disentangle the impact of absorption and source spectrum on the apparent spectrum.

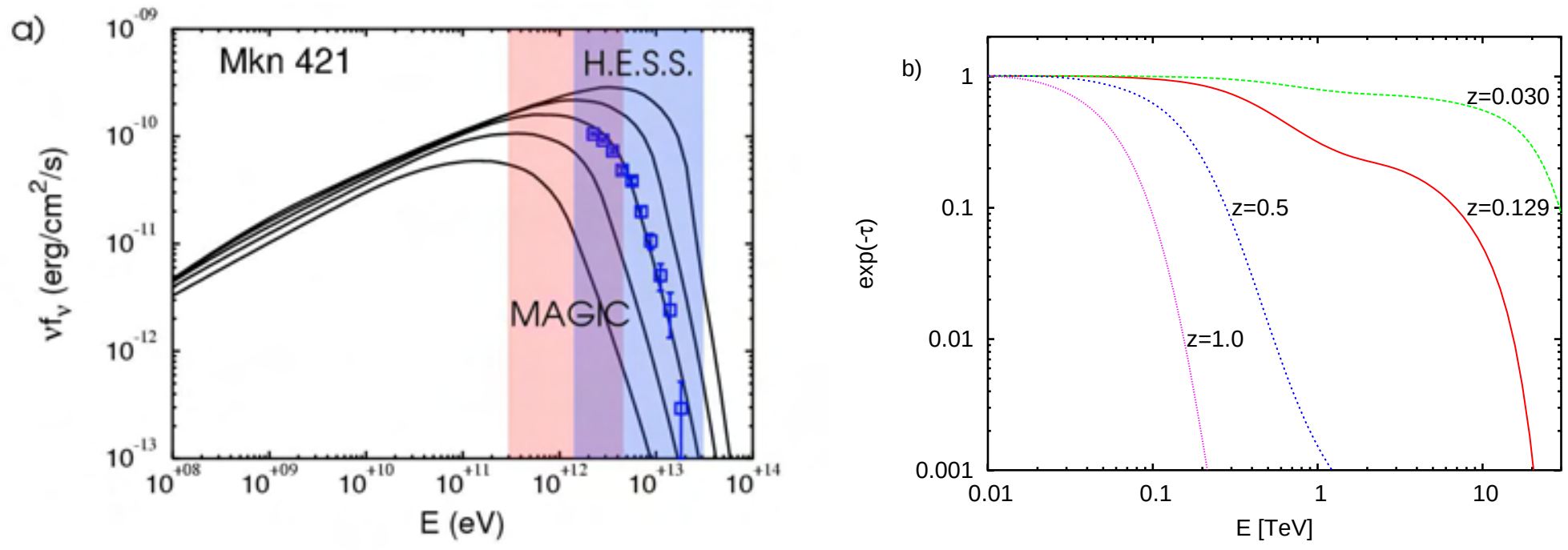

**Figure 3.** a) An example for expected broad band spectral variability of Mkn 421. The energy spectrum is the time-average as measured with the H.E.S.S. telescopes at low elevations in April 2004 [7]. The curves indicate changes of the expected energy spectrum assuming a self-synchrotron Compton model with varying maximum injection energy of the electrons. b) The expected optical depth for different red shifts assuming a model calculation for the extragalactic background light [3].

**Morphology of extended sources at different energies**

It is becoming obvious that many Galactic sources are extended, but this extension may vary with energy [8]. For the interpretation of the acceleration mechanism knowledge of the energy dependent morphology for these objects is necessary. Coordinated (not necessarily simultaneous) effort will be considered to achieve exposure at different energy ranges more efficiently by combining low and high elevation data.

## 4. First results from a joint H.E.S.S. and MAGIC observation

Triggered by an increased activity observed in the X-ray band and by the VERITAS collaboration (H. Krawczynski, priv. comm.) on December 14, 2004 the H.E.S.S. and MAGIC collaborations performed joint observations of Mkn 421. Due to weather conditions common observations were performed only during two nights (December 18 and 19). Moreover, due to observational constrains (such as zenith angle and dark time), the common observational window was only open for 30 minutes on each night. The joint data sample of Mkn 421 encompasses:

1. H.E.S.S. : 1h, zenith angle range = 65-67°, good weather conditions
2. MAGIC : 1h, zenith angle range = 47-54°, good weather conditions.

In addition to these joint observations more data on the Mkn 421 flare were taken by the two experiments without a time overlap. The H.E.S.S. measurements contain 0.5 h at mean zenith angle of 67° on December 16; the MAGIC data contain 2.5 h at mean zenith angle of 15° on December 18 and 19. The results of the analysis of the data [9, 10] show clear detections in the simultaneous data set (6 - 10 $\sigma$) and a reasonable number of photons collected in the different energy bands. The results are encouraging given the comparably low flux state of the source, the short simultaneous observation time, and the non-optimal zenith angles observed. The energy spectrum and a light curve of the common data set are in preparation.

## 5. Concept of a Global Network of Cherenkov Telescopes

With the installation of the four major Cherenkov telescope facilities (CANGAROO III, H.E.S.S. , MAGIC, VERITAS), global networking of these telescope facilities is becoming feasible. Different physics questions could motivate different type of observations (at different longitudes and/or latitudes). Currently the MAGIC and the H.E.S.S. collaborations are organizing a well defined joint observation strategy including fast mutual information exchange on target of opportunity sources.

## Acknowledgements

The authors acknowledge the support and help of both collaborations and the respective teams, funding agencies, and the excellent working conditions on both sites.

# MAGIC observations of high-peaked BL Lacertae objects

M. Meyer[a], K. Mannheim[a], T. Bretz[a], D. Dorner[a], B. Riegel[a], D. Höhne[a] and K. Berger[a] on behalf of the MAGIC collaboration[b]

*(a) Institut für Theoretische Physik und Astrophysik, Universität Würzburg, Am Hubland, 97074 Würzburg, Germany*
*(b) Updated collaborator list at: http://magic.mppmu.mpg.de/collaboration/members/index.html*

Presenter: M. Meyer (meyer@astro.uni-wuerzburg.de), ger-meyer-M-abs1-og23-oral

In a key observation program, the MAGIC Collaboration seeks to find new extragalactic sources at 100 GeV energies by observing bright X-ray selected BL Lacertae objects. Using results of the campaign, we can probe emission models for relativistic jets, the effects of gamma ray attenuation due to interactions with the metagalactic radiation field, and the origin of the extragalactic gamma ray background.

## 1. Introduction

Blazars are radio-loud active galactic nuclei (AGN) viewed at small angles between the jet axis and the line of sight. Due to relativistic beaming they are the brightest and most variable high-energy sources among AGN.

BL Lacertae objects represent a subclass of the blazars, in which broad emission lines are absent or very faint. Apart from the host galaxy's contribution, their spectral energy distribution (SED) is completely dominated by nonthermal emission. Based on the SED, the distinction between low-peaked BL Lacs (LBL), which emit most of their synchrotron power at low frequencies (far-IR, near-IR), and high-peaked BL Lacs (HBL), which emit most of their synchrotron power at higher frequencies (UV-soft-X) was introduced [12].

The observed TeV sources belong to the HBL class, in which the emitted power presumably shows a second peak at very high energies (VHE). An obvious posibility is that, the low-energy peak is due to synchrotron radiation by relativistic electrons, and the high-energy peak is due to inverse-Compton radiation of the same electrons [2]. There could also be two electron populations: a primary population consisting of in situ accelerated electrons, and a secondary one originating in electromagnetic cascades initiated by in situ accelerated protons and nuclei [10].

Despite all the observational differences among the various blazar sub-classes, a unified scheme, in which the luminosity is a fundamental parameter, is consistent with the data [4]. The SEDs of the different blazar types can be considered as a continuous sequence. As shown by [6] this sequence scheme would be consistent with the theoretical synchrotron self compton (SSC) and external compton (EC) models, where the EC component dominates at higher energies. However, the scheme is far from proven, it could also be a mere selection effect. The more luminous sources are more distant (due to redshift evolution), and thus suffer stronger from gamma ray attenuation rendering them invisible at TeV energies. The unified scheme can be tested, for instance, it predicts that sources emitting strongly in the TeV band have a relatively low intrinsic luminosity.

Absorption of $\gamma$-photons by pair production in photon-photon scattering with photons of the Metagalactic Radiation Field (MRF) gives rise to a Gamma Ray Horizon (GRH), which is dependent on the energy of the photons. The cut-off energy, defined as the energy where the optical depth becomes unity, as a function of the redshift of a source has been coined the Fazio-Stecker relation (FSR).

Due to uncertainties in the models of the MRF the FSR shows large differences below redshifts of 0.1, where the TeV $\gamma$-rays are absorbed by scattering off photons from the Cosmic Infrared Backround (CIB), but also for redshifts higher than 1.0, where the attenuation is due to the photons of the diffuse UV radiation field [8]. For sources at redshifts between 0.1 and 1.0, the FSR converges for different models. Up to now, there are only three nearby blazars with published cut-off energies.
Measuring the $\gamma$-ray spectra from blazars at different redshifts allows to test the Fazio-Stecker relation. For nearby blazars, it allows also to distinguish between different models for the CIB.

As a defining property, blazars show time variability on various time scales at all frequencies. In a flaring state, the flux can be an order of magnitude or more higher than in a quiet state. The spectrum generally varies with the flux. At VHE Mkn 421 and Mkn 501 showed flux variability on short time scales, such as the 1996 $\gamma$-ray flare of Mkn 421 with a flux doubling time scale of less than 15 min [5]. Short time variability allows to discriminate between different theoretical models for the dynamic of the jets as well as for the generation of the radiation (SSC, PIC).

## 2. Sample

For the hard X-ray band, a high-sensitivity all-sky-survey does not exist. Therefore we decided to use the blazar hard X-ray compilation from Donato ([3]), which contains 421 continuum spectra from 268 blazars (136 HBL, 63 LBL, 69 Flat Spectrum Radio Quasars (FSRQ)). The list contains blazars with known spectral information in the radio, optical and X-ray band.

### 2.1 Selection criteria

From this 136 HBL objects, a sample of 13 objects is choosen, which satisfies several selection criteria. These criteria should assure, that the object is detectable in less than 20 h, taken into account the current energy threshold of MAGIC (with the dependence on the zenith distance), the attenuation of the $\gamma$-rays by the MRF and the correlation between the X-ray flux and the $\gamma$-ray flux.

1. Redshift
   The first selection criterium is a cut at redshift 0.3. From the FSR one exspects a cut-off energy of about 300 GeV and optical depth of 0.1 at 100 GeV, which means, that the flux at this energy is attenuated by $\sim$20%.

2. X-ray flux
   The different SSC models predict a strong correlation between X-ray and $\gamma$-ray fluxes. Based on the unified scheme we made the assumption for HBL, that the X-ray flux at 1 keV is equal to the gamma-ray flux at 200 GeV. A coarse estimation gives, that a X-ray flux of 2 $\mu$Jy corresponds to a $\gamma$-ray flux of $1.5 \cdot 10^{-11}\mathrm{cm}^{-2}\mathrm{s}^{-1}$ at 200 GeV. We select all objects with a maximum X-ray flux of more than 2 $\mu$Jy.

3. Visibility
   The energy threshold of an IACT depends also on the zenith distance. We require a maximum zenith distance of 30° at the culmination.
   One object was skipped due to a very bright star in the field of view (FOV).

**Table 1.** List of targets with the name of the source (IAU), position, optical brightness, X-ray flux at 1 keV and the observation time of the data taken from August 2004 to June 2005. For Mrk 421 10.3 h are taken in wobble mode.

| source | RA | $\delta$ | $m_V$ | z | $F_x[\mu J]$ | $t_{obs}[h]$ |
|---|---|---|---|---|---|---|
| 0120+340 | 01 23 08.9 | +34 20 50 | 15.2 | 0.272 | 2.4 - 6.3 | - |
| 0317+1834 | 03 19 51.8 | +18 45 35 | 18.1 | 0.190 | 0.2 - 3.1 | 14 |
| 0323+022 | 03 26 14.0 | +02 25 15 | 16.5 | 0.147 | 0.7 - 6.8 | 4 |
| 0414+009 | 04 16 53 | +01 04 54 | 16.4 | 0.287 | 4.6 - 5.2 | - |
| 0806+524 | 08 09 49.2 | +52 18 58 | 14.8 | 0.138 | 4.9 | - |
| 0927+500 | 09 30 37.6 | +49 50 24 | 17.2 | 0.188 | 4.0 | - |
| 1011+496 | 10 15 04.2 | +49 26 01 | 16.2 | 0.200 | 2.2 | - |
| Mrk 421 | 11 04 27.3 | +38 12 32 | 13.5 | 0.031 | 23.9 - 58.4 | 25/10.3 |
| 1218+304 | 12 21 21.9 | +30 10 37 | 16.5 | 0.130 | 7.5 - 10.1 | 10.5 |
| 1415+259 | 14 17 56.6 | +25 43 25 | 16. | 0.237 | 2.9 - 4.3 | 19.9 |
| 1426+428 | 14 28 32.5 | +42 40 25 | 16.5 | 0.129 | 4.6 - 13.4 | 21 |
| Mrk 501 | 16 53 52.2 | +39 45 37 | 13.8 | 0.034 | 8.3 - 46.4 | 36 |
| 1722+119 | 17 25 04.4 | +11 52 16 | 15.8 | 0.018 | 3.6 | 5.7 |
| 1727+502 | 17 28 18.6 | +50 13 11 | 16.0 | 0.055 | 3.6 - 3.7 | - |
| 2344+514 | 23 47 04.9 | +51 42 18 | 14.6 | 0.044 | 3.4 - 6.9 | - |

### 2.2 Targets and observation strategy

The target list contains 15 objects. Four objects, Mrk 421, Mrk 501, H 1426+428 and 1ES 2344+514, are known TeV sources. Eight of the other eleven objects are also suggested by Costamate & Ghisellini to be candidates of TeV emission [1].

The confirmed TeV source 1ES 1959+650 has also been observed by the MAGIC telescope. For this systematic study, it is not included in the target list, due to its higher declination, leading to a culmination at high zenith distance and therefore higher energy threshold. Table 1 contains the list of targets with their position, optical brightness, redshift and the X-ray flux. The visibility of the objects is spread over the whole year. For every object (except for Mrk 421, Mrk 501 and 1ES1959+650, where longer observations are scheduled) a observation time of 20 h is planned. In 20 h, we expect a $5\,\sigma$ detection for the weaker sources even with the more pessimistic flux estimate, while for stronger sources or sources in a flaring state, we expect to unfold spectra and light curves.

## 3. Results

From August to June 2005, nine targets have been observed with a total observation time of $\sim$176 h. For Mrk 421 and 1ES 1959+650 we report a strong detection ([11],[13]). Mrk 501 has been strongly detected, too. As most of the observation took place in the last two months, the analysis of the other targets is still going on. First results will be shown at the conference.

## 4. Acknowledgements

We acknowledge support by the German Federal Ministry of Education and Research (BMBF, 05 CMOMG1/3) and the Astrophysics Institute of the Canary Islands (IAC).

# Status of Silicon Photomultiplier Developments as optical Sensors for MAGIC/EUSO–like Detectors

A. N. Otte[a], B. Dolgoshein[b], H. G. Moser[c], R. Mirzoyan[a], M. Teshima[a]

*(a) Max-Planck-Institut für Physik, Föhringer Ring 6, 80805 Munich, Germany*
*(b) Moscow Engineering and Physics Institute, Kashirskoe Shosse 31, 115409 Moscow, Russia*
*(c) MPI Halbleiterlabor, Otto-Hahn-Ring 6, 81739 Munich, Germany*
Presenter: A.N. Otte (otte@mppmu.mpg.de), ger-otte-N-abs1-og27-oral

A few years ago a new type of photon detector was introduced; the so-called Silicon photomultiplier (SiPM). In this paper we review the working principle of SiPMs and describe the status of our development. Finally, we sketch the possible application of a $5 \times 5\,\text{mm}^2$ SiPM in Air Cherenkov telescopes.

## 1. Introduction

The study of the most violent processes in the universe by future experiments requires photon detectors with photon detection efficiencies (PDEs) higher than that of currently available ones. We are involved in photon detector developments for the MAGIC experiment and future EUSO–like space missions.

MAGIC [2] is currently the world largest air Cherenkov telescope. MAGIC has been constructed to study sources of very high energy $\gamma$–rays (VHE–$\gamma$) in the energy range from a few tens of GeV up to a few tens of TeV. In the experiment, $\gamma$–rays are detected indirectly by collecting Cherenkov photons which are emitted in a VHE–$\gamma$ induced electromagnetic cascade in the atmosphere. For a more comprehensive summary of the physics programm of MAGIC the interested reader is directed to [1].

With an EUSO [4] like experiment, it is planned to study cosmic rays of the highest energies ($\geq 10^{19}$ eV) by detecting from extended air showers fluorescent light with a space born detector, for example as proposed by the EUSO collaboration. The photon yield at the entrance pupil of the detector is in the order of a few hundred to thousand photons distributed over a time window of about $100\,\mu$s.

In both experiments one uses (MAGIC) or plans to use (EUSO) large areas of classical photomultipliers (PMTs) which, beside other disadvantages, exhibit only moderate quantum efficiencies (QE). A light detector with a substantial higher photon detection efficiency ($> 40\%$) would lower the threshold energy of EUSO and MAGIC. This will allow for a considerable overlap with other experiments and thus cross calibration (EUSO/AUGER MAGIC/GLAST) as well as for opening a still unexplored energy region in the electromagnetic spectrum (MAGIC).

We are developing a new type of photon detector — the silicon photomultiplier (SiPM) — which already at the present stage shows a PDE similar to that of PMTs with bialkali photocathodes. In the following sections we will discuss the detector principle and our plans to enhance the photon detection efficiency of these devices.

## 2. SiPM Principle

A SiPM is an array of microcell avalanche photodiodes (APDs) operating in the limited Geiger mode[1]. A photoelectron generated in the depleted region of a Geiger-APD cell is initiating an electrical breakdown which can be easily detected due to the large current flowing. One possible way to quench the breakdown is to use a resistor which is limiting the current through the junction. This so-called passive quenching is applied in SiPMs.

The main disadvantage of single Geiger-APD cells is the inability to distinguish between one or more photoelectrons, as the output signal is solely determined by the capacitance of the diode. This has been overcome some years ago by the idea to combine an array of Geiger-APD cells on the same silicon substrate (e.g. [5, 6, 7]) and interconnect all cells in parallel through integrated quenching resistors (s. Fig. 1). The output signal of the SiPM is the analog sum of all individual cell signals. In applications where the number of photons incident on the SiPM per event is much smaller than the number of cells the output signal is proportional to the number of photons.

[1]We will use the expression Geiger-APD cell for one such APD. In the Geiger mode operation the APDs are reversely biased a few Volts above breakdown.

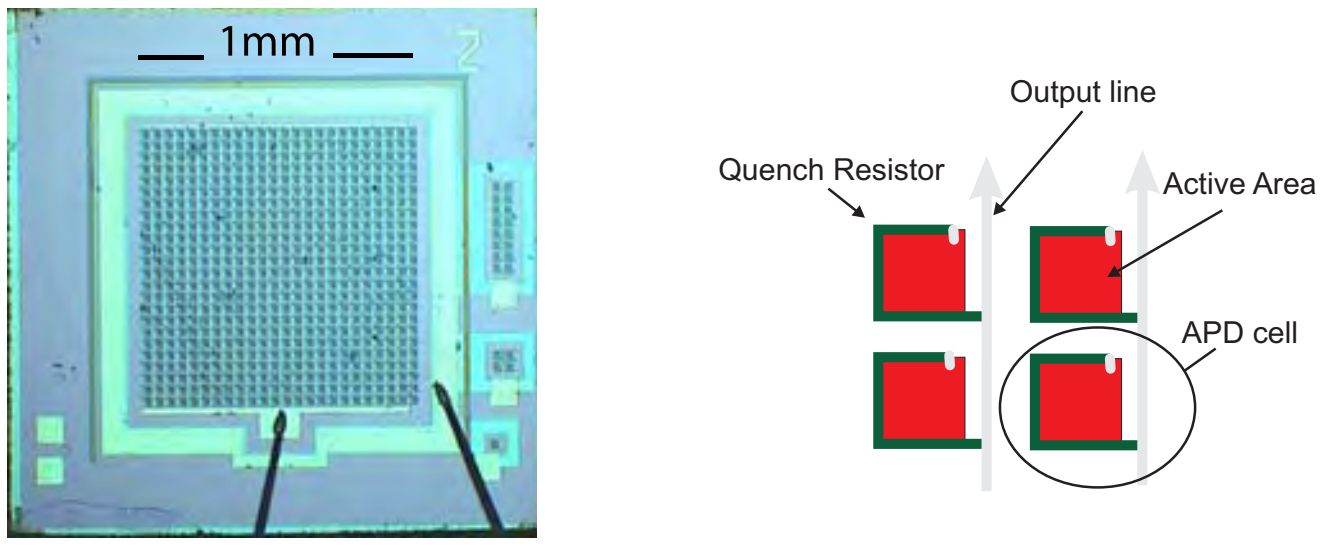

**Figure 1.** The left picture shows a 576 cell SiPM with a total sensor size of $1 \times 1$ mm$^2$. On the right hand side an exemplary view of four Geiger-APD cells is shown which are part of a SiPM (c. f. text for further explaination).

The SiPM is a noisy detector as it is sensitive to every electron generated in or diffusing into the active region. Typical dark rates at room temperature are $10^5 \ldots 10^6$ counts per second and mm$^2$ sensor area. In most applications in astrophysics the dark rate can be sufficiently lowered by moderate cooling to about -50 °C. The aim is to reduce the noise to $\leq 10\%$ of the ubiquitous night sky light background.

The photon detection efficiency of SiPMs depends on many parameters. The most limiting one is the dead space between single cells, which can reach 80% in some devices. Nevertheless it has recently been claimed by some groups that some SiPMs now reach PDEs of 20–30% in the blue wavelength region.

Another interesting feature which is intrinsic to SiPMs is the so-called optical crosstalk between individual cells. This is the result of photons which are emitted during a Geiger breakdown[8] and are migrating to neighboring cells initiating additional cells. This leads to output signals that are too large. There are two solutions for keeping optical crosstalk low. Either the SiPM gain is kept as low as possible or one optically decouples cells from each other. The last solution is currently under investigation by several groups.

The main drawback not to use SiPMs at this stage in astrophysics experiments is due to small sensitive areas. In Table 1 we give an overview of parameters of currently available SiPMs.

**Table 1.** Typical specifications of available SiPM

| **parameter** | **value** |
|---|---|
| Sensor area | $(1 \times 1)$ mm$^2$ $\ldots$ $(5 \times 5)$ mm$^2$ |
| Nr. of Geiger APDs per mm$^2$ | $\sim 100 \ldots \sim 10000$ |
| active area of single SiPM cells[2] | $10\% \ldots 50\%$ |
| peak photon detection efficiencies | $20\% \ldots 30\%$ |
| bias voltage | 30V . . . 100V |
| gain | $10^4 \ldots 10^7$ |
| Geiger APD cell recovery time | $\sim$1 µs |
| typ. noise rate at room temperature | $10^5 \ldots 10^6$ counts/mm$^2$/s |

[2]This includes the area of Geiger APD as well as the space around it which includes the quenching resistor and space to the next neighboring cell

## 3. Ongoing Detector Developments

**Conventional SiPM Design:** In collaboration with MEPhI and Pulsar Enterpise we are developing the above described SiPM concept. Our main efforts are to enhance the PDE in the blue wavelength region beyond 40% and to enlarge the size of SiPMs up to $10 \times 10\,\mathrm{mm}^2$. We try to achieve the first aim by increasing the individual cell size at constant spacing between cells. In this way a geometrical fill factor of 75% seems to be feasible, which might be increased further. An additional increase in light collection can be obtained by using additional light concentrators; this will be explained later in this article.

To reduce optical crosstalk we plan to introduce trenches between cells. First tests have been successfully performed and will now be implemented in the technology of the production of SiPMs [9].

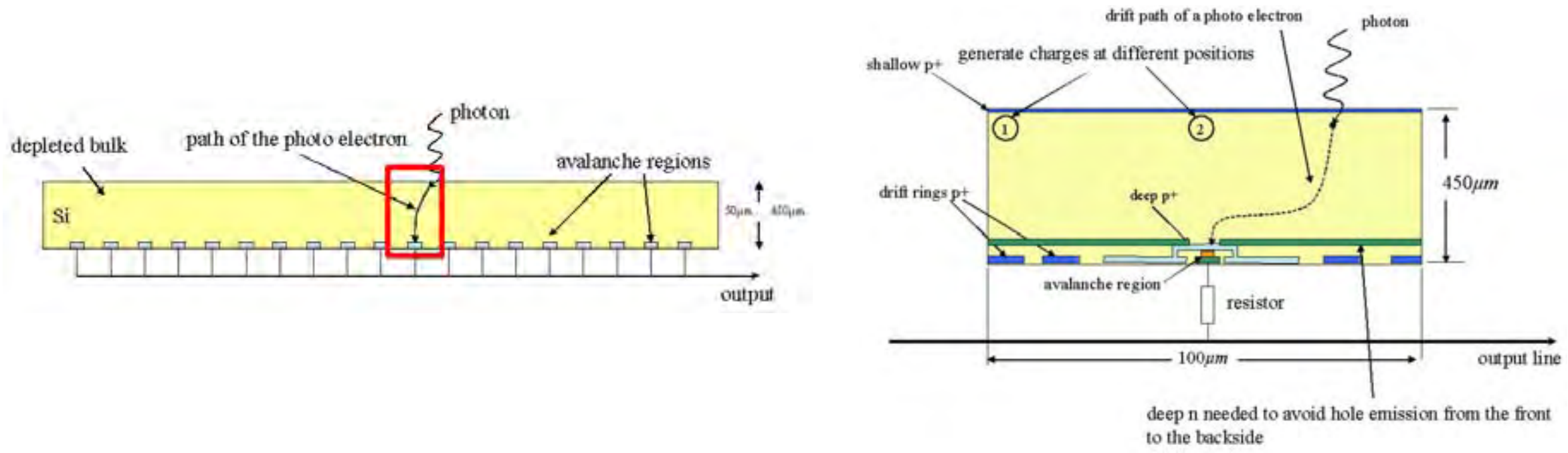

**Figure 2.** Schematic view of the back illuminated SiPM under investigation by the HLL. On the left side, a cross section through a back illuminated SiPM is sketched. On the right side, a blow–up of a single SiPM cell is shown. The drift path of a virtual photo electron from its point of generation until it reaches the Geiger APD is added.

**Backside Illuminated Design** In collaboration with the semiconductor laboratory (HLL) of the Max Planck Institutes for Physics and Extraterrestial Physics we develop a different kind of SiPM which promises to be superior to the above discussed SiPM in terms of the geometrical fill factor [10]. In this new concept the readout node of a silicon drift detector (SDD) is replaced by a Geiger APD. Such a Geiger drift cell (diameter $100\,\mu$m) is the basic cell of a novel back illuminated SiPM (s. Fig. 2). The advantages are small cell capacitances and full area efficiency.

The potential distribution in the cell is constructed such that all photoelectrons generated in the sensitive area are focused with high sensitivity into the Geiger–APD structure [10, 11]. The time jitter in the arrival time of photo electrons at the avalanche region due to their generation at different positions in the cell volume was simulated and found to be smaller than 3 nsec. This is mostly dependent on the geometry of the cell and only in some parts on the drift field. The HLL has a long experience in developing semiconductor detectors with low leakage currents; hence dark rates are expected to be tolerable with some additional moderate cooling. Currently we are translating our simulation results into a technology compatible to the one used by the HLL for other silicon photon and X-ray sensors. In order to verify simulations and evaluate the anticipated parameters of the device we aim to produce test structures within this year.

## 4. Application in Air Cherenkov Telescopes

As SiPM prototypes with sizes of $5 \times 5\,\mathrm{mm}^2$ will be available within the next months, we will discuss shortly a possible application scheme in Air Cherenkov Telescopes, which can also be applied in other experiments which use imaging techniques (e.g. Fluorescence detectors in AUGER).

Practical sizes of picture elements (pixels) in these detectors are typically in the order of a few centimeters. A possible solution to apply $5 \times 5\,\mathrm{mm}^2$ SiPMs will be to compose one pixel out of about 20 SiPMs. The readout scheme of this SiPM matrix pixels is sketched in Figure 3. In this framework, a MMIC is amplifiying each SiPM signal and decoupling the SiPMs. All twenty signals are afterwards combined by an analog adder using a fast current feedback amplifier.

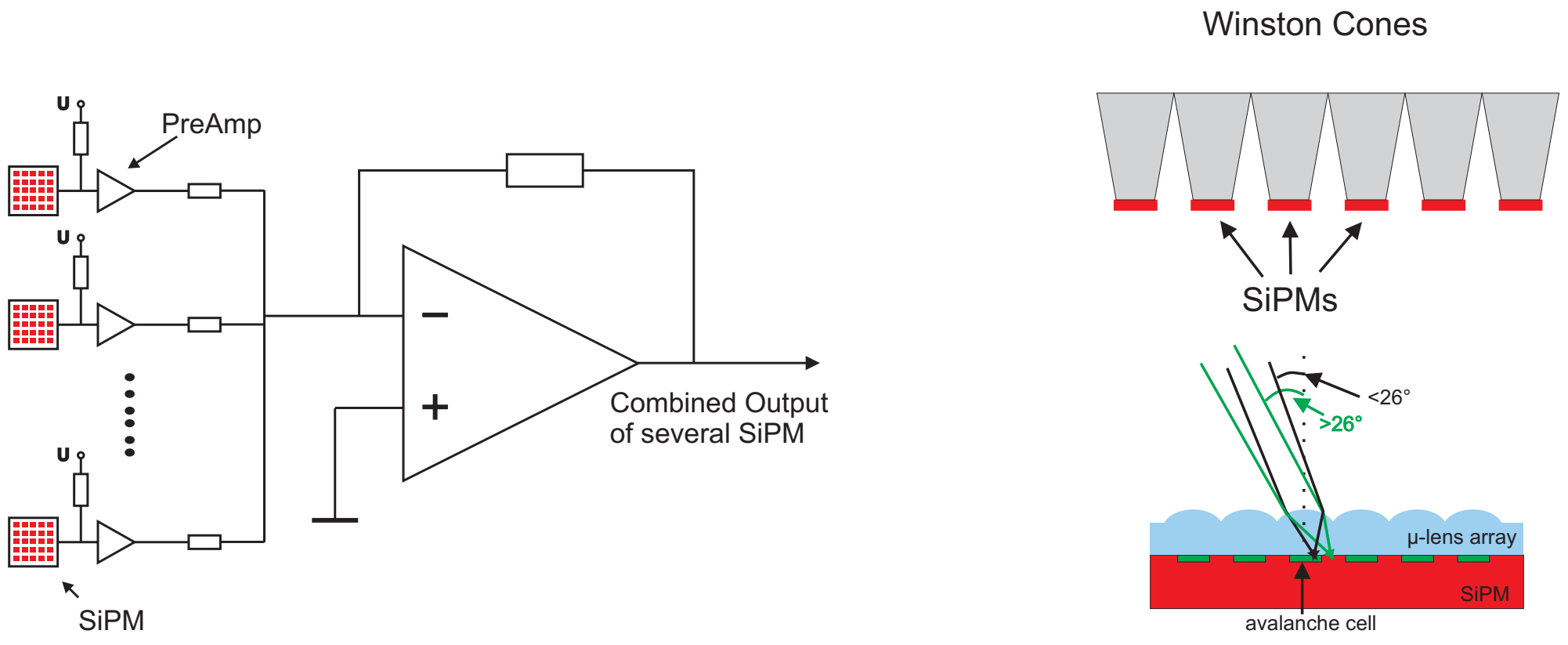

**Figure 3.** Concept of a possible application of a $5 \times 5\,\mathrm{mm}^2$ SiPM in IACTs. On the left side, a readout is sketched which combines several SiPMs to one output channel. On the right side, two possible methods are sketched by which the effective active area can be increased.

There will be a considerable amount of dead space in between individual SiPMs, which has to be compensated. The distance between active areas of single SiPMs has to be about $4\,\mathrm{mm}$ to allow for sufficient space for bonding. Thus, the effective area per SiPM will be $7 \times 7\,\mathrm{mm}^2$, whereas the active area of each SiPM will be $5 \times 5\,\mathrm{mm}^2$, not taking dead space between Geiger cells into account. This factor of two in effective and active area can be recovered by applying non imaging light concentrators to each SiPM. In addition, one can further improve the photon collection efficiency by applying an additional microlens array to each SiPM to focus the incident light into the active area of each Geiger cell as outlined on the right side of Figure 3. This increase in collection area is constrained by the angular acceptance (Liouville theorem).

## 5. Conclusions

We are developing SiPMs for future experiments in high energy astrophysics which need photon detectors with much higher efficiencies in the blue wavelength (UV) region than currently available. The SiPM is a promising replacement candidate for conventional PMTs provided some important development goals will be reached. Firstly, SiPM sizes of at least $5 \times 5\,\mathrm{mm}^2$ have to become available, and secondly, the photon detection efficiency has to be raised above 40%. We try to accomplish both requirements by two independent developments, one together with MEPhI and Pulsar Enterprise, and another at the HLL.

# Study of the MAGIC sensitivity for off-axis observations

J. Rico[a], N. Sidro[a], J. Cortina[a] and E. Oña-Wilhelmi[a] for the MAGIC Collaboration*
[a] *Institut de Fisica d'Altes Energies (IFAE). Universitat Autonoma de Barcelona. 08193 Bellaterra (Barcelona) Spain.*
* *Updated list of collaborators http://wwwmagic.mppmu.mpg.de/collaboration/members/*
Presenter: J. Rico (jrico@ifae.es), spa-rico-J-abs1-og27-poster

We present a study of the sensitivity of the MAGIC telescope for off-axis observations. The results presented have been obtained by the comparison of on- and off-axis Crab Nebula observations, for several positions of the source in the field of view (FOV). The source position is reconstructed for every image using the DISP method. The results are compared with Monte Carlo (MC) simulations. This study allows us to assess the capabilities of MAGIC to perform sky-scans.

## 1. Introduction

The interest of building a catalogue of Very High Energy (VHE) $\gamma$-ray sources have led the different Imaging Air Cherenkov Telescope (IACT) experiments to carry out systematic observations of large portions of the sky (compared to their FOV), mainly along the galactic plane [1, 2]. The recent discovery of 8 new sources using this technique by HESS [2] raises the interest of galactic sky-scans. In this paper we show that MAGIC has the capability to contribute to the growing VHE $\gamma$-ray source catalogue by exploring the part of the galactic plane observable from the Northern Hemisphere.

The critical parameters when performing a sky-scan are the angular resolution and sensitivity of off-axis observations. MAGIC achieves a relatively good angular resolution using the DISP method to reconstruct the direction of arrival on a shower-by-shower basis ($\sigma \sim 0.10^\circ$ at 100 GeV) [3]. In this work we show the sensitivity for off-axis observations evaluated using both MC and real data from Crab Nebula observations. From these studies we derive the sensitivity of MAGIC in view of a future sky scan.

## 2. Data analysis

For the present study we have analyzed data from a set of Crab Nebula observations at low zenith angle ($< 30^\circ$) at 0, 0.3 and 0.4 degrees off-axis angles, respectively. The observations were performed during September 2004 and January 2005. In order to fully characterize the sensitivity up to 1 degree off-axis angle, we have also analyzed a sample of MC simulated $\gamma$-ray events at 0, 0.25, 0.5 0.75 and 1 degrees, and compared the results with those obtained with Crab Nebula observations.

Data are processed using a standard Hillas analysis (for a review, see for example Ref. [4]). The so-called Hillas parameters describe the the shape (WIDTH, LENGTH, SIZE, CONCENTRATION...) and position (DIST, ALPHA...) of the shower image within the camera. The latter parameters are meaningful only when the position of the source within the camera FOV is known. In the present work we do not assume the knowledge of such a position (which is the case e.g. during a sky-scan), and therefore we parameterize the shower images using exclusively the set of variables describing the image shape.

Gamma/hadron separation is performed by means of a Random Forest classification algorithm [5]. A Random Forest is a combination of *tree* predictors such that each tree depends on the values of a random vector sampled independently among the Hillas parameters, and with the same distribution for all trees in the forest. Each tree yields a classification of the shower into a class ($\gamma$-ray or hadron candidate). A new variable called

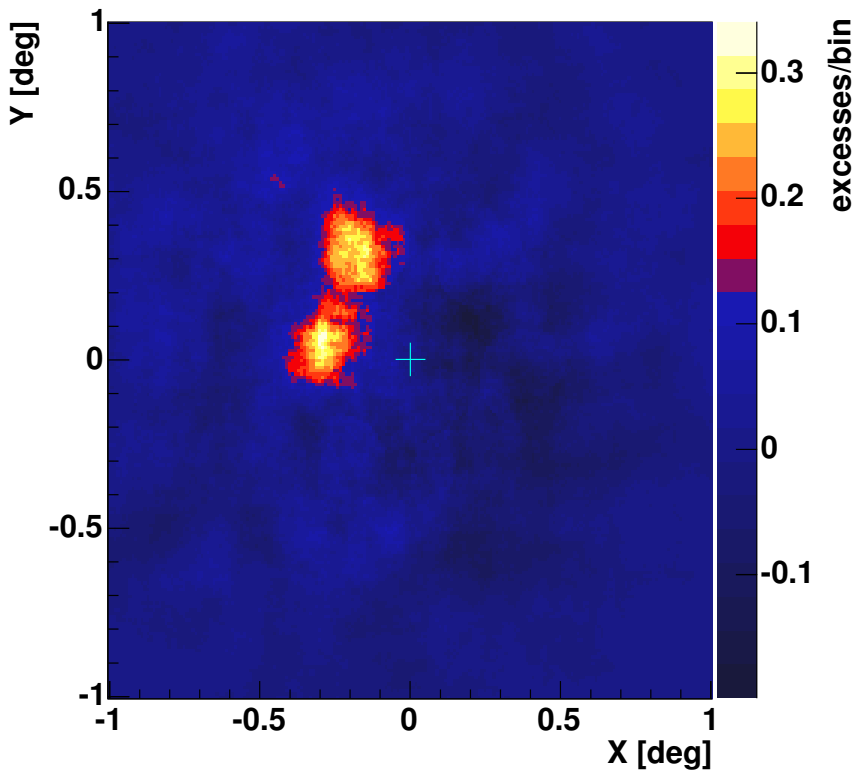

**Figure 1.** Map of excesses of the combined observation of Crab Nebula above 100 GeV at two different positions within the MAGIC camera.

HADRONNESS is computed as the ratio of trees classifying a given image as hadron-like. HADRONNESS is used as the final gamma/hadron discriminator in our analysis.

The direction of arrival of the individual showers is estimated using the DISP method [3]. This method is based on the fact that, for $\gamma$-ray events, the center of gravity of the image and the direction of arrival of the shower within the camera are aligned in the direction of the image maximum elongation. The angular distance between them can be properly parameterized depending only on the image shape (SIZE and WIDTH/LENGTH). Such a parameterization is determined using a sample of MC simulated $\gamma$-ray events and/or a data sample from a well known point-like $\gamma$-ray source (e.g. the Crab Nebula). Using DISP, an angular resolution[1] of $0.10°$ $(0.08°)$ for $E > 100$ $(E > 300)$ GeV is achieved, which allows us to use MAGIC in sky-scans and extended source observations.

For off-axis observations, the background recorded in a given camera position at an angular distance $r$ from the camera center can be computed from the $n$ symmetric positions in the camera, provided that $r\sin(\pi/(n+1)) > 2\sigma$, where $\sigma$ is the angular resolution[2]. For an hexagonal camera as that of MAGIC $n = 1$, 2 or 5. The error on the background estimation is proportional to $n^{-1/2}$, so for each camera position the maximum number of symmetric positions is used. For the region $(r < 2\sigma)$ the background is estimated from the observations of an empty field in the sky, where only background events are expected.

Figure 1 shows the map of the number of excess events above the estimated background for the combined observations of Crab Nebula at two different positions within the MAGIC camera. The angular distance between the positions of the two sources is 0.35 degrees. The sources are clearly resolved.

## 3. Results

We have computed the sensitivity of MAGIC above 100 GeV for off-axis observations as a function of the off-axis angle $(r)$, using both MC $\gamma$-ray and real observation data samples.

[1] We define the angular resolution as the $\sigma$ of a 2-dimensional Gaussian fit to the distribution of reconstructed arrival directions for a point-like source.

[2] Regions suspected to contain signal must be removed during background estimation. The 2-$\sigma$ level ensures that 85% of the signal from a point-like source is removed so that the remaining 15% does not affect background estimation.

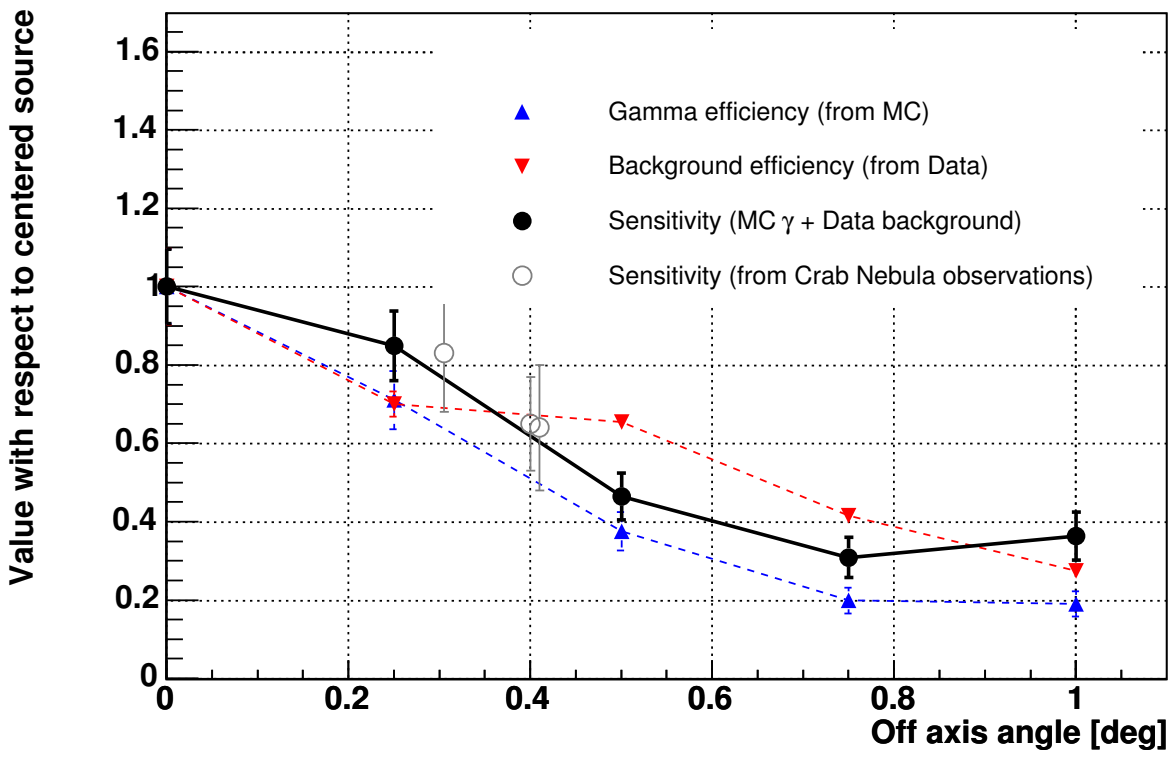

**Figure 2.** Degradation of the sensitivity above 100 GeV together with signal and background efficiencies as a function of the off-axis angle computed both from MC and Crab-Nebula off-axis observations.

MC simulated $\gamma$-rays are used to compute the gamma detection efficiency. For this, about $5 \times 10^5$ gamma-initiated atmospheric showers with a fixed incoming direction were simulated using CORSIKA [6]. The response of MAGIC was simulated using the standard MAGIC reflector and camera MC simulation programs [7], for the different considered observation angles (0, 0.25, 0.5, 0.75 and 1 degrees). The $\gamma$-ray detection efficiency above 100 GeV was compared to that obtained for on-axis observations. The results are shown in figure 2.

Background efficiency was computed using data from observations of an empty FOV at low zenith angle. For the different considered angular distances, the number of background events were compared with that obtained at the camera center. The obtained values were combined with those of the gamma efficiencies to compute the sensitivity degradation: $s_0/s = \frac{N_\gamma/N_\gamma^0}{\sqrt{N_h/N_h^0}}$, where $N_\gamma$ is the number of gamma events for a given off-axis angle, $N_\gamma^0$ the number of gamma events at angle $r = 0$, and similarly for the number of hadrons ($N_h$). The values of the hadron efficiency and sensitivity degradation are shown in figure 2.

In order to check the results obtained with the MC simulation, we have analyzed a set of Crab Nebula observations at 0, 0.3 and 0.4 degrees off-axis angle. We have applied to these data the same analysis as for the MC samples and compared both results (see figure 2). There is a good agreement between the results obtained for MC and data samples.

The integral flux sensitivity[3] of MAGIC above 100 GeV for point-like sources observed on-axis is $2 \times 10^{-11}$ cm$^2$s$^{-1}$ [8]. From this value and those shown in figure 2 we can compute the sensitivity of MAGIC during sky-scans[4]. Figure 3 shows the map of sensitivities of MAGIC during a sky-scan along a given direction (e.g. galactic longitude) for two different values of the angular distance between adjacent observation positions ($d$). In real experimental conditions, the minimum detectable flux depends on the sensitivity and also on the time spent at each observation position. Thus, for a total observation time $T$, covering a total angular distance $D$, the minimum detectable integral flux above 100 GeV (absolute sensitivity of the scan) is given by $F_{\min} = \sqrt{\frac{50D}{Td}}\overline{S}$, where $\overline{S}$ is the sensitivity averaged along the camera. $\overline{S}$ is proportional to $n^{-1/2}$, where $n$ is the number of times a given position in the sky is observed during the scan: $n = 2R/d$ with $R$ the radius of the sensitive area of the camera. Therefore, the absolute average sensitivity of the scan does not depend on $d$, although smaller values are preferred given the greater homogeneity of the sensitivity achieved (see figure 3). For instance, for a scan of a region of $D = 30^\circ$ along the galactic longitude in a total time $T = 300$ hours the absolute sensitivity averaged in the region [-0.5°,0.5°] is $F_{\rm min} = 6.3 \times 10^{-11}$ cm$^2$s$^{-1}$ (about 10% of the Crab-Nebula flux at that energy). The average in the region [-1°,1°] grows up to $9.3 \times 10^{-11}$ cm$^2$s$^{-1}$.

[3] The flux sensitivity is defined here as the minimum integral flux observable with a statistical significance of 5 standard deviations in 50 hours of observation time.

[4] We are considering here a sky-scan in one direction, e.g. the galactic plane. We point the telescope at zero galactic latitude at a given galactic longitude, observe for a given time, and then move the telescope only in galactic longitude.

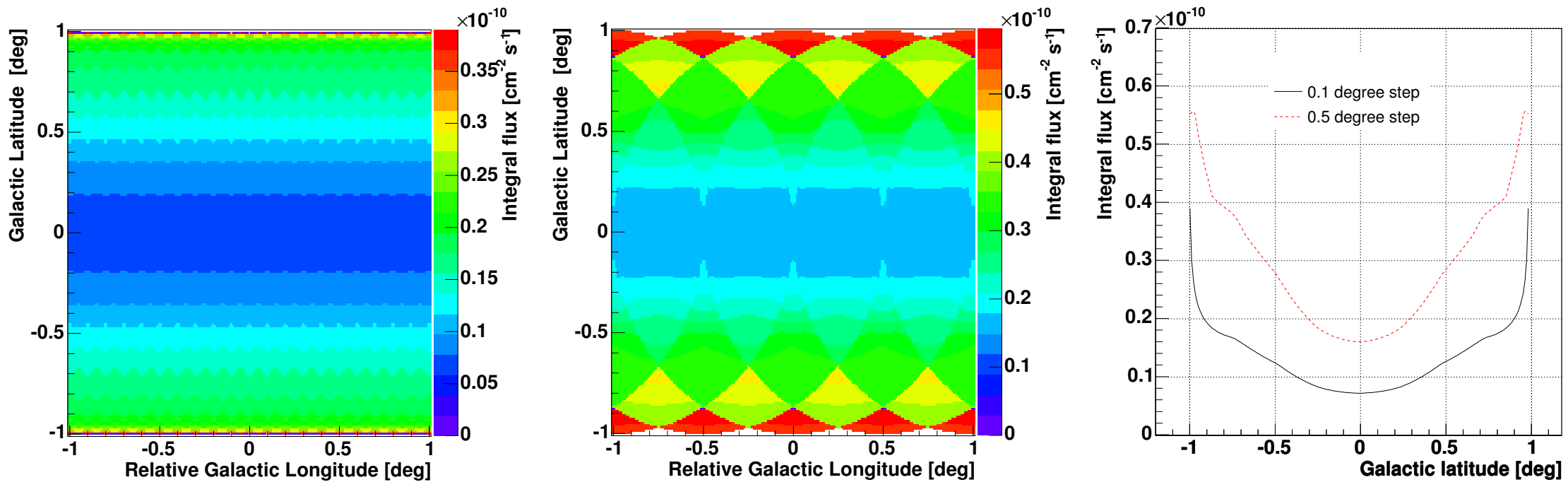

**Figure 3.** Map of sensitivities for point-like sources above 100 GeV achieved by MAGIC for sky-scans with a distance of $0.1^\circ$ (left) and $0.8^\circ$ (center) between adjacent observation positions. Right plot: Average sensitivity as a function of the angular distance to the scan direction for the previously considered conditions.

## 4. Conclusions

The sensitivity of MAGIC for off-axis observations have been assessed by the study of a MC simulated $\gamma$-ray sample and a set of Crab Nebula on- and off-axis observations. The results agree within the statistical errors. We have also computed the MAGIC absolute sensitivity in an eventual sky-scan, obtaining detectable fluxes down to 10% the Crab-Nebula flux above 100 GeV for a 300 hours scan of a 30° length area.

# MAGIC Phase II

C.Baixeras[1], T. Bretz[2], A. Biland[3], R. Bock[4], J. Cortina[5], A. De Angelis[6], D.Ferenc[7], M.V. Fonseca[8], M. Giller[9], F. Goebel[4], M. Hayashida[4], D. Kranich[7], E. Lorenz[4], K. Mannheim[2], M. Mariotti[10], M. Martinez[5], R. Mirzoyan[4], A. Moralejo[10], R. Paoletti[11], F. Pauss[3], N. Pavel[12], L. Peruzzo[10], W. Rhode[13], T. Schweizer[12], K. Shinozaki[4], A. Sillanpaa[15], P. Temnikov[15], M. Teshima[4] and N. Turini[11]

*(1) Universitat Autonoma de Barcelona, Spain*
*(2) Universität Würzburg, Germany*
*(3) Institute for Particle Physics, Swiss Federal Institute of Technology (ETH) Zurich, Swizerland*
*(4) Max-Planck-Institut für Physik, Föhringer Ring 6, 80805 Munich, Germany*
*(5) Institut de Fisica d'Altes Energies, Barcelona, Spain*
*(6) Dipartimento di Fisica dell'Universita di Udine and INFN sez. di Trieste, Italy*
*(7) University of California, Davis, USA*
*(8) Universidad Complutense, Madrid, Spain*
*(9) Division of Experimental Physics, University of Lodz, Poland*
*(10) Dipartimento di Fisica, Universita di Padova and INFN sez. di Padova, Italy*
*(11) Dipartimento di Fisica, Universita di Siena and INFN sez. di Pisa, Italy*
*(12) Institut fur Physik, Humboldt-Universität Berlin, Germany*
*(13) Universität Dortmund, Germany*
*(14) Tuorla Observatory, Piikkiö, Finland*
*(15) Institute for Nuclear Research and Nuclear Energy, Sofia, Bulgaria*
Presenter: M. Teshima (mteshima@mppmu.mpg.de), ger-teshima-M-abs3-og27-oral

MAGIC, the largest ground-based gamma ray telescope in the world, has been in scientific operation since summer 2004. The major motivation of the MAGIC project is to study high energy phenomena in the universe in the unexplored energy region between 10 GeV and 300 GeV. MAGIC-II, a two 17m telescope system with advanced photon detectors and ultra fast readout, is designed to lower the threshold energy further and to simultaneously achieve a higher sensitivity in the stereoscopic / coincidence operational mode. The construction of the second telescope will be completed in 2007. This allows simultaneous observations with the gamma ray satellite missions GLAST and AGILE with a sensitivity not achieved so far by ground-based gamma ray telescopes.

## 1. Introduction

MAGIC was designed back in 1995 with the very clear goal to lower the energy threshold in order to reveal the unexplored energy range between 10 GeV and 300 GeV [1]. The first MAGIC telescope has recently started operation and is now almost reaching its design performance [2]. MAGIC is the first instrument to explore a new energy regime, i.e. >30 GeV energy, with the ground-based imaging Cherenkov technique. We expect that many important physics results will be delivered by this first MAGIC telescope, and our understanding of gamma ray astrophysics in the energy range of a few tens GeV will become deeper. However, we are upgrading MAGIC to MAGIC-II by adding a second telescope and by improving the photon detectors and readout system in order to lower the energy threshold further and, simultaneously, to increase the sensitivity of the telescope. These intentions make sense if we consider the upcoming satellite mission, the Gamma ray Large Area Space Telescope (GLAST) [3]. Simultaneous observations by GLAST and MAGIC-II will provide us with more promising scientific results in the wide energy range of five decades, 100MeV-10TeV, and surely we can deepen our knowledge of the high energy phenomena in the Universe.

The physics objectives of MAGIC-II are widely distributed both in astrophysics and in fundamental physics. **AGNs and GRBs** are prime targets in MAGIC. The function of the fast repositioning (~20 seconds) and the lower threshold energy of 20 GeV will make MAGIC-II the best ground-based detector to study GRBs. Observation at a lower threshold allows us access to larger redshifts, and multiple AGNs up to redshifts ~2 will help to study the **Gamma Ray Horizon** (FSR) determined by the infrared background light, resulting in a better understanding of the cosmological evolution of galaxy formation. The chances for detecting **Quantum gravity** effects using time delays as a function of energy will improve with lowering the threshold, due to an increased number of potential time-variable gamma ray sources, and the larger distances of their positions. Especially GRBs are interesting sources to study this effect. Concerning a search for **Dark matter,** the continuum energy spectrum from neutralino annihilations can only be identified by the measurement with a low threshold.

The GLAST satellite is scheduled for launch in August 2007. Its mission duration will be at least 5 years. GLAST will have an effective collection area of ~ 1 $m^2$ and will be able to detect gamma rays with good energy and direction resolution above 100MeV. MAGIC-II can have a significant overlap with GLAST in the energy region between 20 to 100 GeV. This allows not only the systematic study of high energy sources in the wide energy range, but also the cross calibration of MAGIC and GLAST. GLAST has a wide field of view, which is a strong feature to discover new sources, transient sources like GRBs, and flaring AGNs. MAGIC-II has a large acceptance and provides a better sensitivity for the study of the spectrum and the rapidly changing light curve of AGNs and GRBs in a short time scale. For example, MAGIC-II can supply more precise light curves of high state AGNs after getting an AGN-flare-alert from GLAST. GLAST and MAGIC-II will perfectly complement each other.

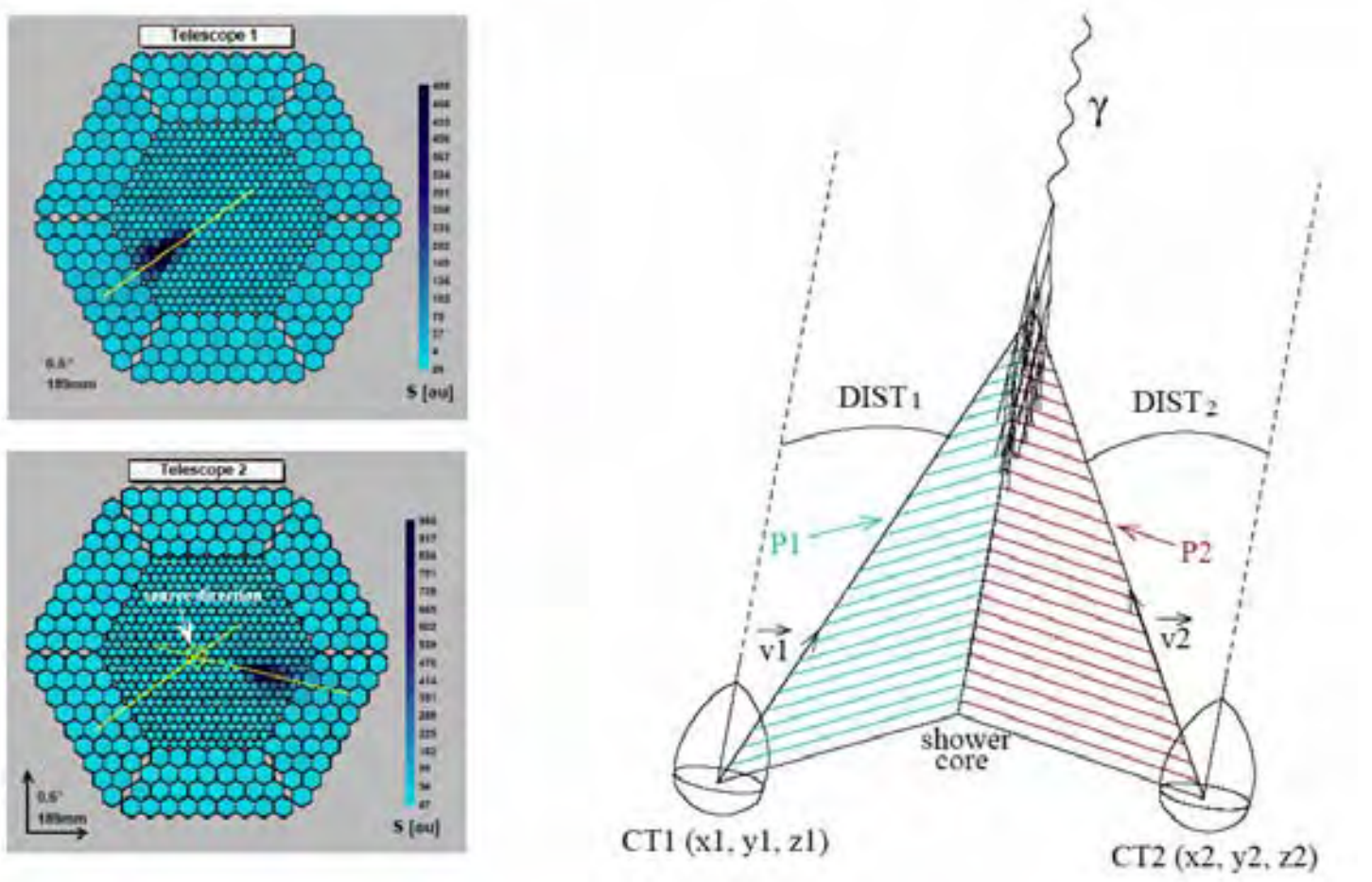

**Figure 1**: Left: reconstruction of the shower direction through the intersection of the major axes of the images in both telescopes. Right: reconstruction of the shower core position assuming that the shower axis is parallel to the pointing direction of the telescopes.

The HESS collaboration plans to upgrade to HESS-II by adding a 28m diameter telescope at the center of the HESS array. This installation will allow to access the energy domain of 10-20 GeV, and HESS-II has a quite similar sensitivity as MAGIC-II. HESS and MAGIC are located in the Southern hemisphere and Northern hemisphere, respectively. Therefore, the entire sky will be covered by low threshold, high sensitivity instruments. In spite of the difference of North and South, the close geographical longitudes of the HESS and MAGIC sites provide a unique opportunity to do a simultaneous observation of interesting variable sources [4]. Irrespective of whether a source is located in the Northern sky or in the Southern sky, one telescope will see it with 10-20 GeV threshold energy, and the other will see it at large zenith angle with high threshold energy but with huge acceptance. We can then study high energy gamma ray emission in the wide energy range between 10 GeV and 10 TeV. The rapidly changing light curve and the time variable spectrum can be studied in a systematic way. Of course, through these simultaneous observations HESS-II and MAGIC-II will be cross-calibrated perfectly.

## 2. MAGIC-II

The second 17m telescope is now under construction at the same site as the first telescope on the Roque de los Muchachos, La Palma, Canary Islands. The second telescope is configured to be situated at a distance of 80m from the first telescope. This distance was optimized after a detailed Monte Carlo simulation. The imaging camera is designed with advanced photon detector HPDs which have a quantum efficiency of 50% around 500nm [5]. The signals will be read out by ultra-fast FADC systems of 2.5 Gsamples/s [6] to reduce background photons from the night sky and to achieve a better gamma / hadron separation by taking into account the time profile of the Cherenkov light. We aim to lower the threshold energy by a factor of two with advanced photon detectors and an ultra fast readout system. The stereo configuration with two telescopes will increase the sensitivity to fainter sources and the quality of the experimental data.

The second telescope substantially increases the sensitivity to gamma rays and also increases the flexibility of the observations. One operational mode is a stereoscopic mode, which allows us to perform deep sky survey. The other is an independent / patrol operational mode, which allows us the fast scanning of different sources. With this mode, we can increase the chance to discover flaring AGNs (AGN patrol).

In the stereoscopic deep observation mode, there are two advantages; a better gamma / hadron separation, and a capability of geometrical reconstruction. The better gamma / hadron separation power will supply us with higher purity gamma ray samples and allow us to study the energy spectra of gamma rays from astronomical objects with higher precision or with less systematic errors. The geometry reconstruction will help to determine energies and arrival directions of individual gamma rays more precisely. The high purity gamma ray samples with geometrical information will also provide good understanding of the characteristics of low energy gamma ray showers experimentally; this will be invaluable information for the conceptual design of the next generation ultimate ground-based gamma ray telescope.

In Figure 2, the expected distributions of the image parameters obtained by Monte Carlo simulation are shown. The $\theta^2$- distribution represents the angular resolution ($\theta$ is the space angle between the source position and the estimated shower direction). The angular resolution, defined as the error circle containing 50% of the events, can be expressed as $\theta_{50\%} \sim 0.20^{o}\ (100\text{GeV/E})^{0.5}$ as a function of gamma ray energies. Mean scaled width and length will be used for gamma / hadron separation. Figure 3 shows the power of event identification. The horizontal axis shows the classification parameter “Hadronness” (0 = pure gamma, 1 = pure hadron). At a hadroness around 0.4, the quality factor reaches the maximum of ~7.

In Figure 4, the sensitivity of MAGIC-II to high energy gamma rays is shown in comparison with MAGIC-I, HESS, VERITAS and GLAST. The sensitivity of MAGIC-II is better by a factor of two than that of MAGIC-I. This means that we need a factor of four less observation time to get the same significance in MAGIC-II as in MAGIC-I, or, in other words, we can see four times more sources in a fixed time period.

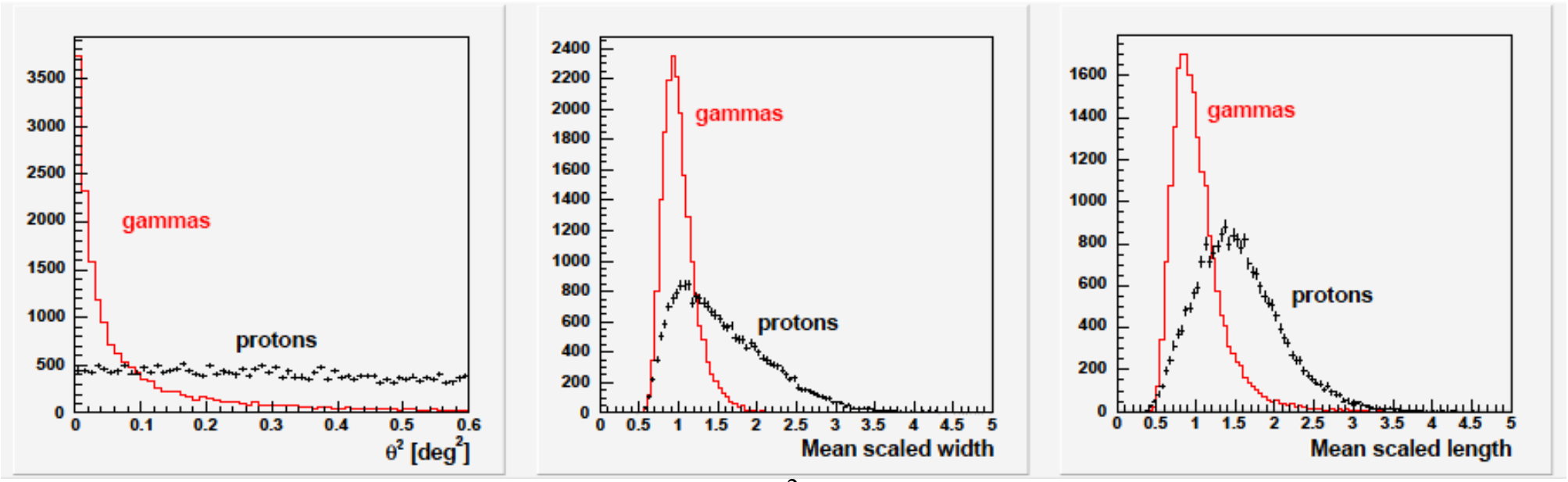

**Figure 2:** The distributions of image parameters, $\theta^2$, mean scaled Width and Length for MAGIC-II.

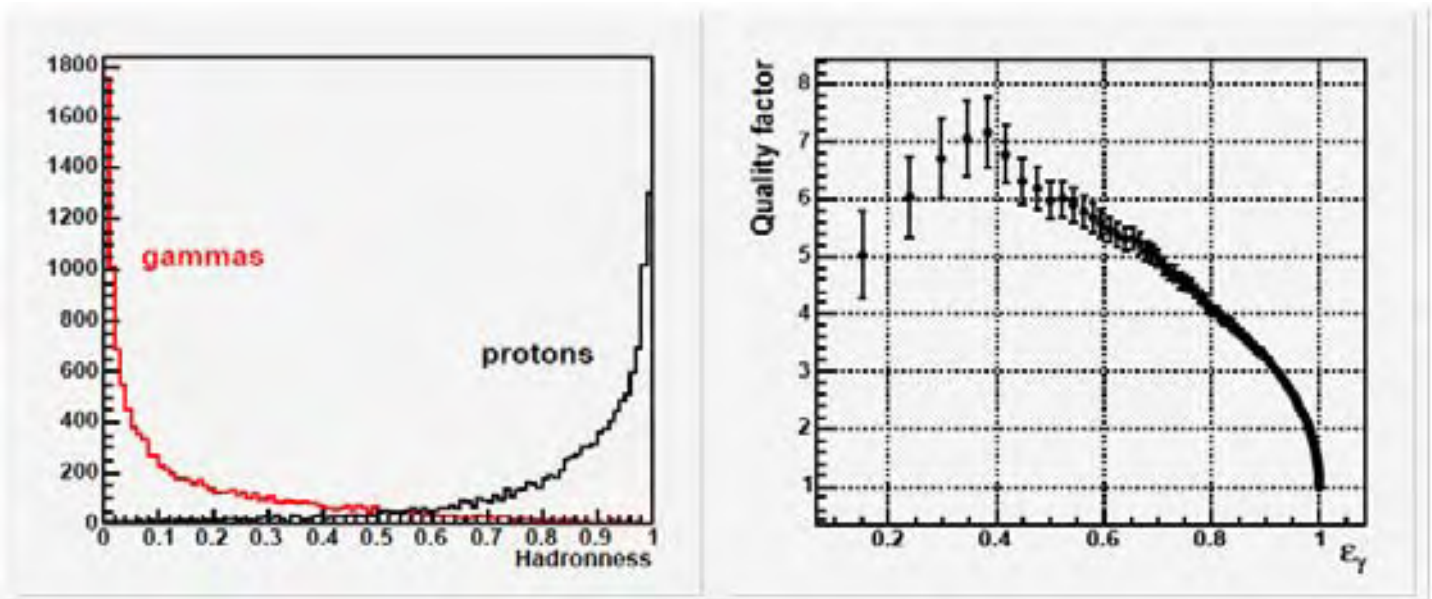

**Figure 3:** Event classification using the Random Forest method in MAGIC-II. Left: hadronness distribution for gammas and protons. Right: quality factor (see text) as a function of the gamma efficiency for different upper cuts in the hadronness.

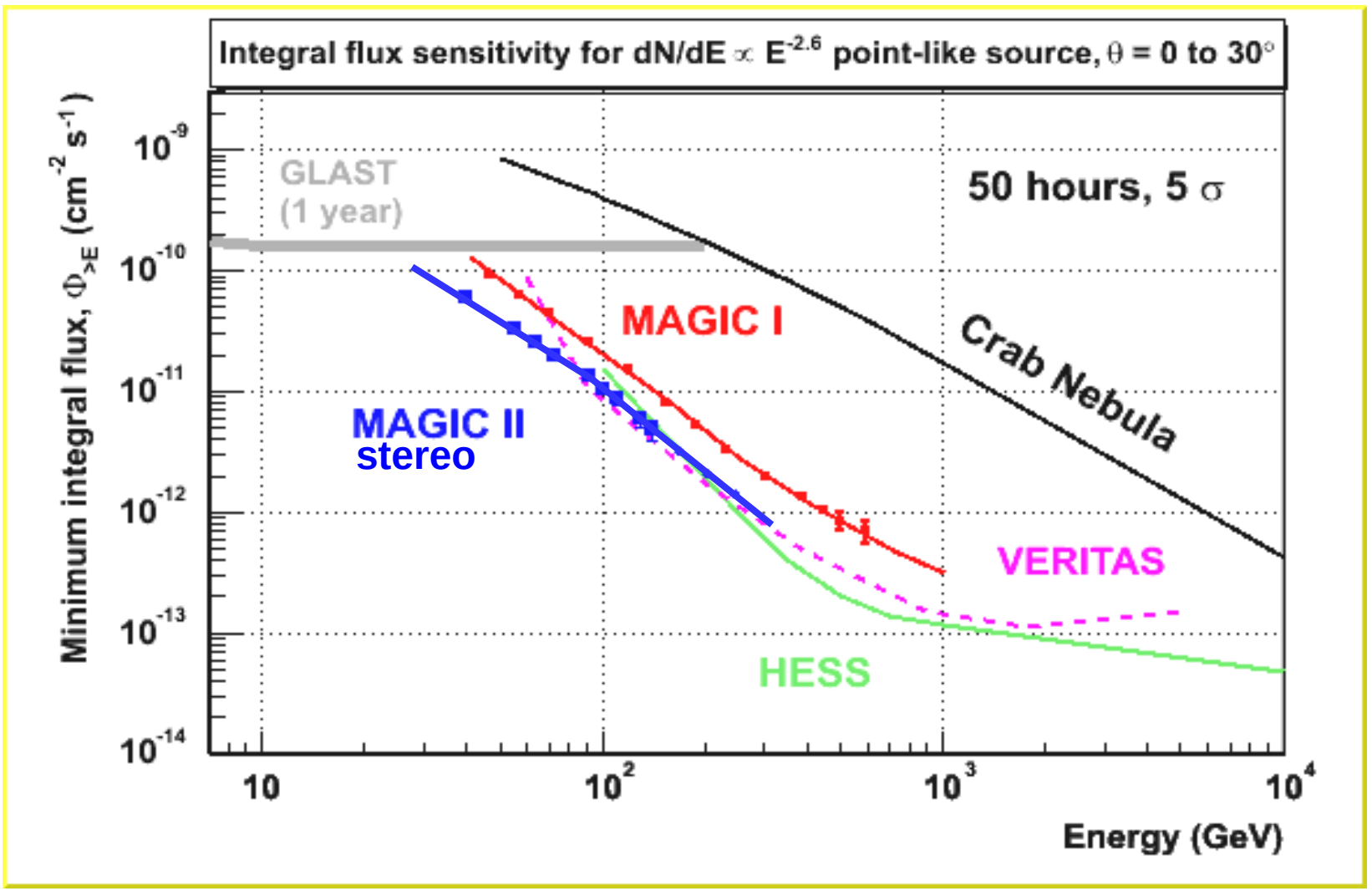

**Figure 4:** Sensitivities of the current MAGIC-I and MAGIC-II (two telescope system with advanced photon detector) are shown in the comparison with GLAST, HESS, and VERITAS. MAGIC-II improves the sensitivity by a factor of two compared to MAGIC-I. Around 30 GeV, the sensitivity of MAGIC-II in 50 hrs will cross the one of GLAST in 1 yr.

## 4. Acknowledgements

The authors thank other collaborators in MAGIC and Dr. M. Altmann for valuable discussions. We would like to thank the IAC for excellent working conditions. The support of the German BMBF and MPG, the Italian INFN and the Spanish CICYT is gratefully acknowledged.